\documentclass[12pt, draftclsnofoot, onecolumn]{IEEEtran}
\usepackage[cmex10]{amsmath}
\usepackage{amssymb}
\usepackage{cite}
\usepackage{graphicx}
\usepackage{array,color}
\usepackage{multirow}
\usepackage{amsmath}
\usepackage{stfloats}
\usepackage{graphicx}
\usepackage{subfigure}
\usepackage{tabularx}
\usepackage{epsfig,epsf,color,balance,cite}
\usepackage{setspace}
\usepackage{bm}
\usepackage{textcomp}
\usepackage[linesnumbered, ruled]{algorithm2e}
\usepackage{subfigure}
\usepackage{caption}
\usepackage{graphicx}
\usepackage{setspace}
\SetKwRepeat{Do}{do}{while}%

\begin{document}

\title{ Joint Power and Blocklength Optimization for URLLC in a Factory Automation Scenario }

\author{
Hong Ren, Cunhua Pan, Yansha Deng, Maged Elkashlan, and Arumugam Nallanathan, \IEEEmembership{Fellow, IEEE}
%\thanks{This work was supported by ...}
\thanks{H. Ren, C. Pan, M. Elkashlan, and A. Nallanathan are with School of Electronic Engineering and Computer Science, Queen Mary University of London, London, E1 4NS, U.K. (Email: {h.ren, c.pan, maged.elkashlan, a.nallanathan}@qmul.ac.uk). Y. Deng is with the Department of Informatics, King's College London, London WC2R 2LS, U.K. (e-mail: yansha.deng@kcl.ac.uk).}
}

\maketitle
\vspace{-1.4cm}
\begin{abstract}
Ultra-reliable and low-latency communication (URLLC) is one of three pillar applications defined in the fifth generation new radio (5G NR), and its research is still in its infancy due to the difficulties in guaranteeing extremely high reliability (say $10^{-9}$ packet loss probability) and  low latency (say 1 ms) simultaneously. In URLLC, short packet transmission is adopted to reduce latency, such that conventional Shannon's capacity formula is no longer applicable, and the achievable data rate in finite blocklength becomes a complex expression with respect to the decoding error probability and the blocklength. To provide URLLC service in a factory automation scenario, we consider that the central controller transmits different packets to a robot and an actuator, where the actuator is located far from the controller, and the robot can move between the controller and the actuator. In this scenario, we consider four fundamental downlink transmission schemes, including orthogonal multiple access (OMA), non-orthogonal multiple access (NOMA), relay-assisted, and cooperative NOMA (C-NOMA) schemes. For all these transmission schemes, we aim for jointly optimizing the blocklength and power allocation to minimize the decoding error probability of the actuator subject to the reliability requirement of the robot, the total energy constraints, as well as the latency constraints. We further develop low-complexity algorithms to address the optimization problems for each transmission scheme. { For the general case with more than two devices, we also develop a low-complexity efficient algorithm for the OMA scheme.} Our results show that the relay-assisted transmission significantly outperforms the OMA scheme, while NOMA scheme performs well when the blocklength is very limited. We further show that the relay-assisted transmission has superior performance over the C-NOMA scheme due to larger feasible region of the former scheme.
\end{abstract}

%\vspace{-0.2cm}
%\begin{keywords}
% URLLC, Mission-Critical IoT, IIoT, 5G NR, MTC.
%\end{keywords}

%
%\newpage
\vspace{-0.4cm}
\section{Introduction}
\vspace{-0.1cm}

The fifth-generation (5G) networks are envisaged to support three pillar use cases: enhanced mobile broadband (eMBB),  massive machine type communication (mMTC), and \emph{mission-critical }internet of things (IoT) \cite{Shafi2017}. Extensive research has focused on eMBB and mMTC, but the research on mission-critical IoT is still in its infancy \cite{Schulz2017,bennis2018ultra,popovski2017ultra,changyangmag,Nielsen2018}. The applications of mission-critical tasks include  factory automation (FA), autonomous driving, remote surgery, smart grid automation, unmanned aerial vehicles (UAVs) control information delivery \cite{mozaffari2018tutorial}, which require ultra reliable and low latency communication (URLLC) \cite{5GTactile,HongTVT,HongVTM}. For example, in Industrial 4.0 \cite{Varghese2014}, wired connection will be replaced by wireless transmission to enhance the flexibility and reduce the infrastructure cost. This change imposes challenging requirements on the wireless transmission in terms of latency and reliability \cite{liangliu2018}. For mission-critical tasks in FA, the transmission duration is expected to be lower than 100 $\mu s$ to allow processing delays during queuing, scheduling, backhaul transmission, and propagation \cite{15iccw}, while guaranteeing the packet error probability of  $10^{-9}$.

In conventional human-to-human (H2H) communications, the transmission delay is relatively long (say 20-30 ms) and the packet size is large (say 1500 bytes), thus  Shannon's capacity can be served as a tight upper bound of the achievable data rate due to the law of large numbers \cite{PetarProceeding}. In contrast, in URLLC, the packet size should be extremely low (say 20 bytes) to support the low-latency transmission \cite{15iccw}. In this case,  Shannon's capacity formula is no longer applicable as the law of large numbers is not valid. Thus, the achievable data rate under short blocklength needs to be retreated. In \cite{Polyanskiy2010}, the achievable data rate in finite blocklength regime has been derived as a complicated function of the signal-to-noise (SNR), the blocklength, and the decoding error probability.

Recently, extensive research attention has been devoted to the short packet transmission (SPT) design \cite{Trillingsgaard2017,changyangshetcom2018,Ostmantcom,yulintvt2016,yulin2016,guspl2018,Lopez2017,prmic2017,xiaoyusun,changyang2018,yulinhu2018}.
In particular, the frame structure is designed in \cite{Trillingsgaard2017} for SPT, where their results showed that it is beneficial to group multiple messages from some users into a single packet based on approximations from finite blocklength information theory. In \cite{changyangshetcom2018}, She \emph{et al.} studied the network available range maximization problem by dynamically selecting the transmission modes between device-to-device (D2D) and cellular links. The non-asymptotic upper and lower bounds on the coding rate for
SPT over a Rician memoryless block-fading channel were derived in \cite{Ostmantcom} under a given packet error probability requirement. The overall error probability of relay-assisted transmission under finite blocklength was derived in \cite{yulintvt2016} under the assumption of perfect channel state information (CSI). They further extended this model to the quasi-static Rayleigh channels where only the average CSI is available at the source in \cite{yulin2016}, as well as to the two-way amplify-and-forward relay network in \cite{guspl2018}. Recently, the delay and decoding error probability were analyzed in \cite{Lopez2017} for simultaneous wireless information and power transfer (SWIPT) relay-assisted system, where the relay first harvests energy from the source and then uses the harvested energy to forward the source's information to the destination node.

The aforementioned studies \cite{Trillingsgaard2017,changyangshetcom2018,Ostmantcom,yulintvt2016,yulin2016,guspl2018,Lopez2017} mainly focused on the performance analysis of finite blocklength transmission. In order to design a practical URLLC system, it is imperative to intelligently optimize the resource allocation including blocklength and power allocation under the given error probability and latency requirements. Unfortunately, the achievable coding rate expression is neither convex nor concave with respect to the blocklength and the transmit power, which brings the difficulty in obtaining the globally optimal solution\cite{changyangmag}. This motivates the recent studies in resource allocation for the SPT in \cite{prmic2017,xiaoyusun,changyang2018,yulinhu2018,pan2019joint}. Specifically, the average throughput and the max-min throughput optimization under the latency constraint was solved via the exhaustive search method with high complexity in \cite{prmic2017}. Sun \emph{et al.} in \cite{xiaoyusun} considered the SPT for a two-user downlink non-orthogonal multiple access (NOMA) system, with an aim to maximize the throughput of user 1 subject to the throughput requirements for user 2. Note that the decoding error probability requirement has not been considered in \cite{prmic2017} and \cite{xiaoyusun}, and the throughput is less important in URRLC as only control signals or measurement data with small packet size are transmitted in URLLC. In \cite{changyang2018},  She \emph{et al.} jointly optimized the uplink and downlink transmission blocklengths to minimize the required total bandiwidth based on statistical channel state information (CSI). However, the optimization is based on the simplified expression of  the rate for SPT, which cannot accurately characterize the relationship between the decoding error probability and  blocklength. In addition, several approximations are involved in the derivation of the decoding error probability for each user due to the fact that only statistical CSI is available. Most recently, Hu \emph{et al.} \cite{yulinhu2018} considered SWIPT in relay-assisted URLLC systems, where the SWIPT parameters and blocklength are jointly optimized to maximize the reliability performance. However, the decoding error probability at the relay cannot be guaranteed and the power is assumed to be fixed in \cite{yulinhu2018}. Most recently, in \cite{pan2019joint} we jointly optimize the blocklength and unmanned aerial vehicle's (UAV's) location to minimize the decoding error probability  while guaranteeing the latency requirement and decoding error probability target. However, the power allocation was not considered. Furthermore, the optimization over UAV's location is obtained by observing the curve of the second-order derivative of the objective function over location variable without strict proof.

In this paper, we consider a typical mission-critical scenario (i.e., a FA scenario), where the central controller needs to transmit a certain amount of different data to two devices within a given transmission time and under a very low packet error probability. One device named actuator is located far away from the controller, while the other device named robot can move between the controller and the actuator. We consider four fundamental transmission schemes, namely,  orthogonal multiple access (OMA), NOMA, relay-assisted transmission and cooperative NOMA (C-NOMA). In this scenario, we aim for jointly optimizing the blocklength and the transmit power of these two devices to minimize the decoding error probability for the actuator while guaranteeing the decoding error probability for the robot, taking into account the energy and blocklength constraints, which were not considered in \cite{xiaoyusun,changyang2018,yulinhu2018} and new methods needs to be developed. The main contributions of this paper are summarized as follows:
\begin{enumerate}
  \item For the OMA scheme, we first prove that both the decoding error probability and energy constraints hold with equality at the optimal point, and then propose a novel iterative algorithm to obtain  tight lower and upper bounds of the blocklength to reduce the search complexity. A low-complexity algorithm is proposed to find the globally optimal solution of transmit power. { For the case of more than two devices, we also develop a novel low-complexity algorithms to find the suboptimal solution of the optimization problem.}
  \item For the NOMA scheme, the search set of blocklength is first derived to reduce the search complexity. In contrast to the OMA case, the decoding error probability function for each given blocklength in the NOMA case is non-continuous with respect to the transmit power, which complicates the optimization problem. Fortunately, we rigorously proved that the decoding error probability holds with equality at the optimal point, such that  the one-dimensional line search algorithm can be used to find the optimal solution. We also provide a sufficient condition when the decoding error probability function is a convex function, which facilitates the application of a low-complexity bisection search method.
  \item For the relay-assisted scheme, we also adopt the iterative algorithm to reduce the search complexity of blocklength. Unlike the OMA and NOMA schemes, the decoding error probability constraint of relay-assisted transmission does not hold with equality. To resolve this issue, we fix the blocklength, such that the original optimization problem is reduced to a one-dimension search optimization problem.
  \item For the C-NOMA scheme, we adapt the iterative algorithm to reduce the search complexity of blocklength, and then one-dimension search is proposed to find the optimal transmit power. For the special case, low-complexity bisection search method is applied.
  \item To compare the performance of our proposed four transmission schemes, we perform extensive simulation results, which show that the relay-assisted scheme significantly outperforms the other three schemes for most times in terms of both the decoding error probability and the network availability. Our results demonstrate the effectiveness of relaying transmission in enhancing the reliability performance in the industrial automation scenario.
\end{enumerate}

The remainder of this paper is organized as follows. In Section \ref{systemmodel}, the system model and the problem formulation are provided. In Section \ref{scheme}, the transmission scheme is presented. { The general case with more than two devices is considered in Section \ref{jfejoref}.}  Simulation results and analysis are presented in Section \ref{simulation}. Finally, Section \ref{conclusion} concludes the paper.

\IEEEpeerreviewmaketitle

\vspace{-0.35cm}
\section{System Model}\label{systemmodel}
\vspace{-0.3cm}
\subsection{System model}
\begin{figure}
\centering
%  % Requires \usepackage{graphicx}
\includegraphics[width=3.1in]{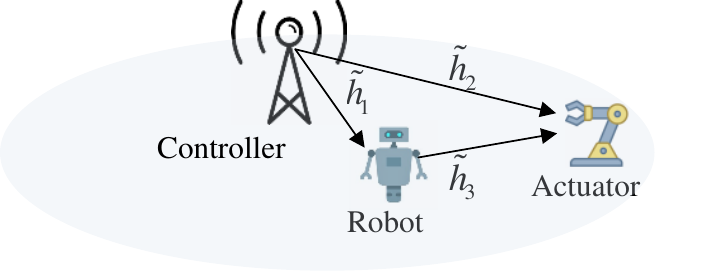}
\vspace{-0.1cm}
\caption{Illustration of a Factory Automation Scenario.}
\vspace{-1cm}
\label{systemodel}
\end{figure}

\vspace{-0.1cm}
Consider a downlink communication in one factory, where a central controller serves a robot and an actuator as shown in Fig. \ref{systemodel}. The robot is assumed to be located in the vicinity of the controller, and the actuator is far away from the controller. Both the robot and the actuator are equipped with a single antenna. The controller needs to transmit two small packets to the two devices. The packet sizes for the actuator and the robot are assumed to be the same, and are denoted as $D$ bits.

The transmission of these two packets is subject to a latency constraint, i.e., the transmission has to finish within $M$ symbols or channel uses. The transmission time corresponds to $t_{\max}=MT_s$ seconds, where $T_s$ is the symbol duration that is equal to $1/B$ with $B$ as the system bandwidth. For the applications with URLLC requirement, short frame structure is adopted and the end-to-end delay should be kept within 1 ms \cite{5GTactile}, which  is much shorter than the channel coherence time. Hence, the channels are quasi-static fading and remain constant during the whole transmission. The channel fading coefficients from the central controller to the robot and the actuator are denoted as ${{\tilde h}_1}$ and ${{\tilde h}_2}$, respectively.  The channel fading coefficient between the robot and the actuator is denoted as ${{\tilde h}_3}$.  We also assume that these channels are perfectly known at the controller, and the total energy consumption of the system should be below ${\tilde E_{{\rm{tot}}}}$ Joule.  Since we have assumed that the actuator is far away from the controller, the channel power gain $\left| {{{\tilde h}_2}} \right|^2$ is very small.

\vspace{-0.25cm}
\subsection{Achievable data rate for a simple point-to-point system}
\vspace{-0.1cm}
The data rate (coding rate) $R$ of a communication system is defined as the fraction of the number of information bits to the number of transmission symbols.  According to  Shannon's coding theorem, the Shannon capacity is defined as the highest coding rate that there exists an encoder/decoder pair whose decoding error probability  becomes negligible when the blocklength approaches infinity \cite{shannon2001mathematical}. However, in URLLC, the blocklength for each frame is limited and small, in this case, the decoding error probability at the receiver cannot be ignored.

 In URLLC scenarios, the required transmission delay is much shorter than the channel coherence time, thus the channel is quasi-static. According to the results in \cite{weiyang}, for a simple point-to-point communication system transmitting over a quasi-static Rayleigh fading channel, the channel dispersion is zero and the achievable  data rate converges to the outage capacity as the blocklength increases. However, the closed-form expression of the outage capacity for short-packet transmission is unavailable. In \cite{Polyanskiy2010}, the normal approximation was adopted to approximate the coding rate $R$ at finite blocklength, which is given by
\vspace{-0.2cm}
\begin{equation}\label{jfoeih}
  R \approx {\log _2}(1 + \gamma ) - \sqrt {\frac{V}{m}} \frac{{{Q^{ - 1}}\left( \varepsilon  \right)}}{{\ln 2}},\vspace{-0.2cm}
\end{equation}
where $m$ is the channel blocklength, $\varepsilon $ is the decoding error probability, $\gamma$ denotes the signal-to-noise ratio (SNR) at the receiver, ${Q^{ - 1}}\left(  \cdot \right)$ is the inverse function $Q(x) \!=\! \frac{1}{{\sqrt {2\pi } }}\int_x^\infty  {{e^{ - \frac{{{t^2}}}{2}}}dt} $, and $V$ is given by $V \!=\! 1 - {(1 + \gamma )^{ - 2}}$. As shown in the numerical results in \cite{weiyang}, this approximation is very accurate when $m$ is larger than 50, which is the case in our simulations.  {From (\ref{jfoeih}),
the decoding error probability can be obtained as follows:
\begin{equation}\label{dwfaref}
  { \varepsilon }=Q\left( {f\left( {{{\gamma}},{m},D} \right)} \right),
\end{equation}
where $f\left( {\gamma ,m,D} \right)= \ln 2\sqrt {\frac{m}{V}} \left( {{{\log }_2}(1 + \gamma ) - \frac{D}{m}} \right)$.} In the following, we aim to jointly optimize the transmission blocklength and power to minimize the decoding error probability for four different transmission schemes.

\vspace{-0.3cm}
\section{Transmission schemes}
\vspace{-0.1cm}
\label{scheme}
In this section, we aim for designing efficient resource allocation algorithms to minimize the decoding error probability of the actuator under three sets of constraints: 1) the packets for robot and actuator need to be transmitted within $M$ symbols; 2) the robot should satisfy its reliability requirement; 3) the total consumed energy should be kept within ${\tilde E_{{\rm{tot}}}}$. The OMA,  NOMA, relay-assisted transmission, and C-NOMA transmission schemes are studied in the following subsections.

\vspace{-0.6cm}
\subsection{OMA transmission}
\vspace{-0.15cm}
The OMA scheme is the simplest transmission scheme, where the controller serves the robot and the actuator in two different orthogonal channel uses or blocklengths. In detail,   the controller transmits signal $x_1$ to the robot with $m_1$ blocklength.  Due to this orthogonal property,  the received signal at the robot can be represented as
\vspace{-0.75cm}
\begin{spacing}{1.2}
\begin{equation}\label{rejih}
\vspace{-0.15cm}
{y_{1}} = \sqrt {{p_1}} {{\tilde h}_1}{x_1} +  {n_1},
\end{equation}
\end{spacing}
\noindent where $p_1$ is the transmit power of the robot, $n_1$ is the zero-mean additive complex white Gaussian noise (AWGN) with variance $\sigma _1^2$, $x_1$ carries information knowledge for the robot with packet size $D$. Hence, the coding rate at the robot is given by $D/m_1$.

  From (\ref{rejih}),  the received signal to noise ratio (SNR) at the robot is given by
\vspace{-0.25cm}
\begin{spacing}{1.0}
\begin{equation}\label{dewrfjroi}
\vspace{-0.1cm}
{\gamma _1} = {p_1}{h_1},
\end{equation}
\end{spacing}
\noindent where $h_1=|\tilde{h}_1|^2/{\sigma _1^2}$  denotes the normalized channel gain from the controller to the robot. Then, according to (\ref{dwfaref}), the decoding error probability of $x_1$ at the robot  is given by
\vspace{-0.2cm}
\begin{spacing}{1.0}
\begin{equation}\label{dwdewd}
\vspace{-0.1cm}
{\varepsilon _1} =  Q\left( {f\left( {{{\gamma_1}},{m_1},D} \right)} \right).
\end{equation}
\end{spacing}

The controller transmits signal $x_2$ to the actuator with blocklength equal to $m_2$. The corresponding error probability at the actuator is derived as
\vspace{-0.4cm}
\begin{equation}\label{dwddewdwed}
\vspace{-0.1cm}
{\varepsilon _2} =  Q\left( {f\left( {{{\gamma_2}},{m_2},D} \right)} \right), \vspace{-0.3cm}
\end{equation}
where ${\gamma _2} = {p_2}{h_2}$ with $p_2$ as the transmit power of the actuator and $h_2=| {\tilde h}_2|^2 /{\sigma _2^2}$ as the normalized channel gain for the actuator. Without loss of generality (w.l.o.g.), we assume that in this paper the robot has higher normalized channel gain than the actuator, i.e., ${{h}_1}>{{ h}_2}$.

The resource allocation problem for the OMA transmission can be formulated as:
\vspace{-0.7cm}
\begin{spacing}{1.0}
\begin{subequations}\label{initial-pro1}
\begin{align}
\vspace{-0.1cm}
\mathop {\min }\limits_{\left\{ {{m_1},{m_2},{p_1},{p_2}} \right\}} \;\;\;&  {{ \varepsilon }_2}\\
{\rm{s.t.}}\;\;\;& {{ \varepsilon }_1} \le \varepsilon _1^{\max },\label{rgegrtgue}\\
&{m_1}{p_1} + {m_2}{p_2} \le {E_{{\rm{tot}}}},\label{fgrtgrtihre}\\
&m_1+m_2 \le  M,\label{freocdsdi}\\
& m_1, m_2\in\mathbb{Z},\label{ofrrefetypo}
\end{align}
\end{subequations}
\end{spacing}
\noindent where constraint in (\ref{rgegrtgue}) is the decoding error probability requirement of the robot, constraint (\ref{fgrtgrtihre}) ensures the system total energy consumption is within a budget $\tilde{E}_{\rm{tot}}=E_{\text{tot}}T_s$, (\ref{freocdsdi}) is the constraint on the latency constraint, and constraint (\ref{ofrrefetypo}) ensures that the blocklength for each transmission phase is integer with $\mathbb{Z}$ denoting the positive integer set.  The maximum decoding error probability $\varepsilon _1^{\max }$ is assumed to be much less than 0.1 to ensure the stringent reliability requirement, and this assumption holds for the remaining transmission schemes. As a result, $\varepsilon _1$ should be smaller than 0.1. Then, the inequality $\frac{D}{m_1}< {\rm{log}}_2(1+\gamma_1)$ should hold.

To solve the optimization problem in (\ref{initial-pro1}), we first provide the following lemma.

\emph{\textbf{Lemma 1}}: Constraints (\ref{rgegrtgue}) and (\ref{fgrtgrtihre}) hold with equality at the optimum solution.

\emph{Proof}: \upshape Please see Appendix \ref{lemma1}.  \hfill\rule{2.7mm}{2.7mm}

With $m_1$ and $m_2$ to be integers, the exhaustive search method can be used to find the optimal solution. To reduce the search complexity when $M$ is large, we shorten the search range of $m_1$ and $m_2$.  In the following, we aim to derive the bounds of $m_1$ and $m_2$.

\subsubsection{The upper and lower bounds of $m_1$ and $m_2$}

Since $\varepsilon _1^{\max }$ is assumed to be a very small value that is much smaller than $10^{-1}$, a necessary condition for constraint (\ref{rgegrtgue}) to hold is that $ {\log _2}\left( {1 + {p_1}{h_1}} \right) > D/{{m_1}}$ \footnote{  This can be proved as follows: $\varepsilon_1= Q\left( {f\left( {\gamma_1,m_1,D} \right)} \right)<\varepsilon _1^{\max }<0.5=Q\left( 0 \right)$. Since Q-function is a decreasing function, we have ${f\left( {\gamma_1,m_1,D} \right)}>0$. By substituting the expression of $f\left( {\gamma_1,m_1,D} \right)$, the proof is complete.}, which leads to
\begin{spacing}{1.1}
\vspace{-0.3cm}
\begin{equation}\label{joihjtj}
\vspace{-0.1cm}
 {p_1} > \left({{2^{\frac{D}{{{m_1}}}}} - 1}\right)/{{h_1}}.
\end{equation}
\end{spacing}
On the other hand, based on the energy constraint (\ref{fgrtgrtihre}), we have: $ p_1<E_{{\rm{tot}}}/m_1$. Thus, the blocklength allocation of the robot $m_1$ should satisfy the following inequality:
\begin{spacing}{1.2}
\vspace{-0.2cm}
\begin{equation}\label{dewjfoireh}
\vspace{-0.1cm}
  {E_{{\rm{tot}}}} > \frac{{{m_1}}}{{{h_1}}}({2^{\frac{D}{{{m_1}}}}} - 1) \buildrel \Delta \over = g({m_1}).
\end{equation}
\end{spacing}
To investigate the properties of $g(m_1)$, the first-order and second-order derivatives of function $g({m_1})$ w.r.t. $m_1$  are give by
\vspace{-1.1cm}
\begin{spacing}{1.1}
\begin{eqnarray}
\vspace{-0.4cm}
g'({m_1}) &=& {2^{\frac{D}{{{m_1}}}}} - 1 - \ln 2 \cdot \frac{D}{{{m_1}}}{2^{\frac{D}{{{m_1}}}}},\\
\vspace{-0.4cm}
g''({m_1}) &=& {\left( {\ln 2} \right)^2} \cdot \frac{{{D^2}}}{{m_1^3}}{2^{\frac{D}{{{m_1}}}}} \ge 0.\label{deri_gm}
\vspace{-0.1cm}
\end{eqnarray}
\end{spacing}
Thus, $g'({m_1})$ is a monotonically increasing function of $m_1$, and we have
\vspace{-0.3cm}
\begin{equation}\label{frehiuy}
\vspace{-0.1cm}
  g'({m_1}) \le \mathop {\lim }\limits_{{m_1} \to  + \infty } g'({m_1}) = 0.\vspace{-0.2cm}
\end{equation}
 Hence, function $g({m_1})$ is a monotonically decreasing function of $m_1$. Then, we can find the lower bound of $m_1$ that satisfies the inequality (\ref{dewjfoireh}) which is denoted as $m_1^{\rm{lb}(0)}$, and $m_1$ should be no smaller than $m_1^{\rm{lb}(0)}$, i.e., ${m_1} \ge m_1^{{\rm{lb}}({\rm{0}})}$. {Similarly, for practical applications, the decoding error probability of the actuator ${ \varepsilon }_2$ should be very small, e.g., much lower than 0.5. In this case, the inequality $ {\log _2}\left( {1 + {p_2}{h_2}} \right) > D/{{m_2}}$ should hold, which leads to
\vspace{-0.2cm}
\begin{equation}\label{kopqw}
  {p_2} > \left({{2^{\frac{D}{{{m_2}}}}} - 1}\right)/{{h_2}}.
\end{equation}}
By using the inequality $m_2\le M-m_1^{\rm{lb}(0)}$, we have
\begin{equation}\label{jotrjgho}
  {m_2}{p_2} \ge \frac{{{m_2}}}{{{h_2}}}\left( {{2^{\frac{D}{{{m_2}}}}} - 1} \right) \ge \frac{{M - m_1^{{\rm{lb}}({\rm{0}})}}}{{{h_2}}}\left( {{2^{\frac{D}{{M - m_1^{{\rm{lb}}({\rm{0}})}}}}} - 1} \right) \buildrel \Delta \over = {A^{(0)}}.
\end{equation}
By using constraint (\ref{fgrtgrtihre}), we have
\vspace{-0.2cm}
\begin{equation}\label{koitjhotj}
  {E_{{\rm{tot}}}} - {A^{(0)}} > \frac{{{m_1}}}{{{h_1}}}\left( {{2^{\frac{D}{{{m_1}}}}} - 1} \right).
\end{equation}
By using (\ref{koitjhotj}), the updated lower bound of $m_1$ can be obtained, and denoted as $m_1^{\rm{lb}(1)}$. Similar to  (\ref{jotrjgho}), we can obtain $A^{(1)}$ by substituting $m_1^{\rm{lb}(1)}$ into $m_1^{\rm{lb}(0)}$, and the lower bound of $m_1$ can be obtained by using (\ref{koitjhotj}), where $A^{(0)}$ is replaced by $A^{(1)}$, and the updated lower bound is denoted as $m_1^{\rm{lb}(2)}$. Repeat the above procedure until $m_1^{{\rm{lb}}(n)}=m_1^{{\rm{lb}}(n-1)}$ or $m_1^{{\rm{lb}}(n)}=M-m_2^{{\rm{lb}}(0)}$. Then, denote the final lower bound of $m_1$ as $m_1^{\rm{lb}}$.

 This procedure is proved to converge as follows. By using ${A^{(0)}} \ge 0$ and comparing (\ref{dewjfoireh}) and (\ref{koitjhotj}), we can obtain $m_1^{\rm{lb}(1)}\ge m_1^{\rm{lb}(0)}$, and thus ${A^{(1)}} \ge A^{(0)}$, which leads to $m_1^{\rm{lb}(2)}\ge m_1^{\rm{lb}(1)}$. Hence, the sequence of the lower bound $m_1^{\rm{lb}(n)}$ is monotonically increasing. Furthermore, the sequence is upper bounded by $M-m_2^{\rm{lb}(0)}$. As a result, the sequence generated by the above iterative procedure is guaranteed to converge.

 By using the similar iterative procedure, we can also obtain the lower bound of $m_2$, which is denoted as $m_2^{\rm{lb}}$. As a result, the search region of $m_1$ is given by $m_1^{\rm{lb}}\le m_1\le (M-m_2^{\rm{lb}})\buildrel \Delta \over = m_1^{\rm{ub}} $.

For each given $m_1$, we need to find the search range of $m_2$, which is detailed as follows. The optimal $p_1$ can be obtained by solving the equation ${\varepsilon _1} = \varepsilon _1^{\max }$ with given $m_1$, which is denoted as $p_1^*$. The solution can be readily obtained by using the bisection search method due to the fact that ${\varepsilon _1}(p_1)$ is a monotonically decreasing function of $p_1$.  Then, we have
\vspace{-0.2cm}
\begin{spacing}{1.2}
\begin{equation}\label{jmohjuhhu}
{E_{{\rm{tot}}}} - {m_1}p_1^* = {m_2}{p_2} \ge \frac{{{m_2}}}{{{h_2}}}\left( {{2^{\frac{D}{{{m_2}}}}} - 1} \right).
\end{equation}
\end{spacing}
Hence, the lower bound of $m_2$ with given $m_1$ (denoted as $m_2^{\rm{lb}}(m_1)$) can be obtained from (\ref{jmohjuhhu}), which is the minimum integer that satisfies (\ref{jmohjuhhu}). Obviously, the upper bound of $m_2$ with given $m_1$ is $M-m_1$. Hence, the search region of $m_2$ is given by $m_2^{\rm{lb}}(m_1)\le m_2\le (M-m_1)$.

\subsubsection{Algorithm to solve Problem (\ref{initial-pro1})}

Based on the above analysis, the algorithm to solve Problem (\ref{initial-pro1}) is given in Algorithm \ref{hgirjoj}.
The main idea can be summarized as follows. For each given integer value of $m_1$ that satisfies  $m_1^{\rm{lb}}\le m_1\le m_1^{\rm{ub}} $, we calculate the value of ${{ \varepsilon }_1}$ when $p_1$ is set as ${E_{{\rm{tot}}}}/m_1$. If ${{ \varepsilon }_1}>\varepsilon _1^{\max }$, then the value of $m_1$ is not feasible, and we increase the value of $m_1$ by one and continue to check the updated $m_1$. Otherwise, we apply the bisection search method to find the value of $p_1$ such that ${{ \varepsilon }_1} = \varepsilon _1^{\max }$ due to the monotonically decreasing property of decoding error probability ${{ \varepsilon }_1}$ w.r.t. $p_1$ \cite{xiaoyusun}. By using Lemma 1, we have $m_2p_2=E_{{\rm{tot}}}-m_1p_1$. The search range of $m_2$ is given by $m_2^{\rm{lb}}(m_1)\le m_2\le M-m_1$. For each given $m_2$, the corresponding $p_2$ is given by $p_2=(E_{{\rm{tot}}}-m_1p_1)/m_2$, and we can calculate the value of ${ \varepsilon }_2$. For each feasible $m_1$, we can find the optimal solutions for $m_2$ and $p_2$ that yield the minimum value of ${ \varepsilon }_2$, respectively. At last, we check all feasible   $m_1$ in the range of $m_1^{\rm{lb}}\le m_1\le m_1^{\rm{ub}} $, and choose the final globally optimal solution.

\begin{algorithm}
\setstretch{0.4}
  \caption{Algorithm for Problem (\ref{initial-pro1}) }\label{hgirjoj}
  \SetKwInOut{Input}{Input}\SetKwInOut{Output}{Output}
 \Input  { $h_1, h_2, D, M, \varepsilon _1^{\max }, E_{{\rm{tot}}}$}
  \Output  {$p_1^\star, p_2^\star, m_1^\star, m_2^\star$}
  Apply the iterative procedure to calculate $m_1^{\rm{lb}}, m_1^{\rm{ub}}$ and $m_2^{\rm{lb}}$;

  \For { $m_1=m_1^{\rm{lb}}: m_1^{\rm{ub}}$ }
  {Set $p_1={E_{{\rm{tot}}}}/m_1$, and calculate the value of ${{ \varepsilon }_1}$.

  \eIf{${{ \varepsilon }_1}>\varepsilon _1^{\max }$}
  {The current $m_1$ is not feasible, and return to the next $m_1$;}
  {Use (\ref{jmohjuhhu}) to find the lower bound of $m_2$, denoted as $m_2^{\rm{lb}}(m_1)$. Apply the bisection search method to find the value of $p_1$ such that ${{ \varepsilon }_1} = \varepsilon _1^{\max}$;

  \For{$m_2=m_2^{\rm{lb}}(m_1): M-m_1$}
  {Calculate $p_2=(E_{{\rm{tot}}}-m_1p_1)/m_2$, and the value of ${ \varepsilon }_2$, denoted as ${ \varepsilon }_2(m_1, m_2)$.}
  Given $m_1$, find the blocklength $m_2$ with the minimum value of ${ \varepsilon }_2(m_1, m_2)$:
  \[{\left. {m_2^\#} \right|_{{m_1}}} = \mathop {\arg \min }\limits_{m_2^{{\rm{lb}}} \le {m_2} \le M - {m_1}} {\varepsilon _2}\left( {{m_1},{m_2}} \right).\]
  }
  }
  Return $m_1^ \star  = \mathop {\arg \min }\limits_{m_1^{{\rm{lb}}} \le {m_1} \le m_1^{{\rm{ub}}}} {\varepsilon _2}\left( {{m_1},{{\left. {m_2^\#} \right|}_{{m_1}}}} \right),m_2^ \star  = {\left. {m_2^\#} \right|_{m_1^ \star }}$ and the corresponding $p_1^\star$ and $p_2^\star$.
\end{algorithm}

\subsubsection{Special case of Problem (\ref{initial-pro1})}
In steps  8-10 of Algorithm \ref{hgirjoj}, one has to calculate the value of  ${ \varepsilon }_2$ for each $m_2$, which may incur high complexity.  In this subsection, we consider one special case when
the SNR value $\gamma$ is very high, i.e., $\gamma\gg 1$. In this case, $V$ in (\ref{dwfaref}) can be approximated as one, i.e., $V \approx 1$ \footnote{ In general, when $\gamma> 20$  dB, the value of $V$ is larger than 0.99, which can be approximated as one.}. The optimization problem in this special case can be efficiently solved. Specifically, the decoding error probability in (\ref{dwfaref}) can be approximated as
\begin{equation}\label{sfreref}
   \tilde \varepsilon =Q\left( {{\tilde f}\left( {{{\gamma}},{m},D} \right)} \right),
\end{equation}
where ${\tilde f}\left( {\gamma ,m,D} \right) = \ln 2\sqrt {m} \left( {{{\log }_2}(1 + \gamma ) - \frac{D}{m}} \right)$.

For given $m_1$ and $p_1$, the product of $m_2$ and $p_2$ should satisfy ${m_2}{p_2} = {E_{{\rm{tot}}}} - {m_1}{p_1} \buildrel \Delta \over = {E_2}$ according to Lemma 1. Then,   the original problem defined in  (\ref{initial-pro1}) can be transformed to the following optimization problem:
\vspace{-0.5cm}
\begin{spacing}{1.3}
\begin{equation}
\mathop {\min }\limits_{m_{\rm{2}}^{{\rm{lb}}} \le {m_2} \le M - {m_1},{m_2} \in \mathbb{Z}} Q\left( {\tilde f\left( {{\gamma _2},{m_2},D} \right)} \right).
\end{equation}
\end{spacing}
Since $Q$-function is a decreasing function, the above problem is equivalent to the following problem by substituting $p_2=E_2/m_2$ into it as
\vspace{-0.4cm}
\begin{spacing}{1.2}
\begin{equation}\label{dewfre}
\!\!\mathop {\max }\limits_{m_{\rm{2}}^{{\rm{lb}}} \le {m_2} \le M \!-\! {m_1},{m_2} \in \mathbb{Z}} \ln 2\sqrt {{m_2}}\! \left(\! {{{\log }_2}\!\left( \!{1 \!+\! \frac{{{E_2}{h_2}}}{{{m_2}}}} \!\right) \!\!-\!\! \frac{D}{{{m_2}}}}\! \right).
\end{equation}
\end{spacing}
To solve the above problem, we first relax the integer variable $m_2$ to a continuous variable, and define
\vspace{-0.8cm}
\begin{spacing}{1.2}
\begin{equation}\label{dewafe}
\tilde g({m_2}) \buildrel \Delta \over = \sqrt {{m_2}} \left( {{{\log }_2}\left( {1 + \frac{{{E_2}{h_2}}}{{{m_2}}}} \right) - \frac{D}{{{m_2}}}} \right).
\end{equation}
\end{spacing}
In the following theorem, we provide a sufficient condition for $\tilde g({m_2})$ to be a concave function.

\emph{\textbf{Theorem 1}}:  $\tilde g({m_2})$ is a concave function when $\frac{{{E_2}{h_2}}}{{M - {m_1}}} \ge e - 1$, where $e$ is the natural constant.

\emph{Proof}: \upshape Please see Appendix \ref{theorem1}.  \hfill\rule{2.7mm}{2.7mm}

When the condition in Theorem 1 is satisfied, Problem (\ref{dewfre}) is a convex optimization problem.
If $\tilde g'(m_{\rm{2}}^{{\rm{lb}}})\le 0$, the optimal $m_2$ is given by $m_2=m_{\rm{2}}^{{\rm{lb}}}$. If $\tilde g'(M - {m_1})\ge  0$, the optimal $m_2$ is $m_2=M - {m_1}$. Otherwise, the optimal $m_2^*$ satisfies  $\tilde g'(m_{\rm{2}})= 0$, and the low-complexity bisection search method can be used to find $m_2^*$. The final optimal integer $m_2$ is the one with lower objective value for its two neighbor integers, i.e.,  $\left\lfloor {m_2^*} \right\rfloor $ and $\left\lfloor {m_2^*} \right\rfloor +1$.

\vspace{-0.1cm}
\subsection{NOMA transmission}\label{NOAMscheme}
\vspace{-0.1cm}

In NOMA transmission, superposition coding is employed at the controller so that the controller can transmit signals to the two devices simultaneously with different power levels. {The controller allocates higher transmit power to the user with lower channel gains and lower power to the one with higher channel gains. On the one hand, the robot decodes the actuator's signal first. If decoding correctly, the robot will subtract the actuator's signal from its received signals and decodes its own signal. This is the so-called successive interference cancellation (SIC). Otherwise, it has to decode its own signal by treating actuator's information as interference.  On the other hand, the actuator directly decodes its own signal by treating the robot's signal as interference since the controller allocates higher transmit  power than the robot.  To implement this scheme, it is crucial that the robot  knows whether SIC is successful or not. To this end, we assume that the controller sends the actuator's channel coding information along with the robot's channel coding information to the robot through dedicated error-free channels. The channel coding information for both devices are different and the channel coding can assist in detecting whether the decoded information is correct or not. Hence, the robot knows whether the SIC is successful or not. In general, the channel coding information changes when CSI changes, which is much longer than the URLLC transmission. Hence, each channel coherence time can accommodate multiple URLLC transmissions. Then, the coding information only needs to be transmitted to the robot at the beginning of channel coherence time, which  causes negligible overhead consumption.}

 In NOMA, the transmission blocklength for two devices is equal to $M$.
Specifically, the received signals at the robot and the actuator are given by
\vspace{-0.25cm}
\begin{spacing}{1.2}
\begin{equation}\label{dusdwefru}
\begin{array}{l}
{y_{1}} = \sqrt {{p_1}} {{\tilde h}_1}{x_1} + \sqrt {{p_2}} {{\tilde h}_1}{x_2} + {n_1},\\
{y_{2}} = \sqrt {{p_1}} {{\tilde h}_2}{x_1} + \sqrt {{p_2}} {{\tilde h}_2}{x_2} + {n_2},
\end{array}
\end{equation}
\end{spacing}
\noindent where the notations in (\ref{dusdwefru}) has the same meaning as those in the OMA transmission scheme. For the robot, it  first decodes the actuator's signal, where the decoding signal to interference plus noise ratio (SINR) is given by
\begin{spacing}{1.2}
\vspace{-0.7cm}
\begin{equation}\label{denbbvf}
 \gamma _2^1 = \frac{{{p_2}{h_1}}}{{{p_1}{h_1} + 1}}.
\end{equation}
\end{spacing}
Following (\ref{dwfaref}), the decoding error probability of $x_2$ at the robot can be written as $\varepsilon _2^1 =Q\left( {f\left( {{{\gamma_2^1}},{M},D} \right)} \right)$. This equivalently indicates that the information $x_2$ can be accurately cancelled at the robot with probability $1-\varepsilon _2^1$. Note that this is different from the infinite blocklength case in NOMA, where perfect decoding  can be achieved by the robot. If the SIC is successful, the robot decodes its signal $x_1$ by removing the decoded signal $x_2$. By using the first equality in (\ref{dusdwefru}), the SINR of decoding the signal $x_1$ is given by
 \vspace{-0.35cm}
 \begin{spacing}{1.0}
 \begin{equation}\label{dewkikooewo}
 {\gamma _1} = {p_1}{h_1}.
 \end{equation}
 \end{spacing}
 Thus, following (\ref{dwfaref}), the decoding error probability of $x_1$ at the robot under perfect SIC condition is given by ${\varepsilon _1} =Q\left( {f\left( {{{\gamma_1}},{M},D} \right)} \right)$.
 However, if the SIC fails, the robot will decode its information $x_1$ while treating $x_2$ as interference, and the corresponding SINR is given by
 \vspace{-0.25cm}
\begin{equation}\label{dertthr}
\vspace{-0.2cm}
  {{\hat \gamma_1 }} = \frac{{{p_1}{h_1}}}{{{p_2}{h_1} + 1}}.
\end{equation}
Thus, the decoding error probability of $x_1$ at the robot is given by ${{\hat \varepsilon_1 }}= Q\left( {f\left( {{{\hat\gamma_1}},{M},D} \right)} \right) $. Based on the above discussion, the decoding error probability of $x_1$ at the robot is Bernoulli-distributed. With probability $1 - \varepsilon _2^1$, the decoding error probability is equal to  ${\varepsilon _1}$, and with probability $\varepsilon _2^1$, it is equal to ${{\hat\varepsilon_1}}$. Hence, the average decoding error probability of $x_1$ at the robot is formulated as
\vspace{-0.3cm}
\begin{spacing}{1.15}
\begin{equation}\label{detrgrihrf}
\vspace{-0.2cm}
  {{\bar \varepsilon }_1} = {\varepsilon _1}(1 - \varepsilon _2^1) + {{\hat\varepsilon_1}}\varepsilon _2^1.
\end{equation}
\end{spacing}

Recall that the actuator directly decodes its own signal by treating the signal from the robot as interference, and its SINR is given by
\vspace{-0.5cm}
\begin{spacing}{1.2}
\begin{equation}\label{edwgtrgtrew}
  {{\gamma_2}} = \frac{{{p_2}{h_2}}}{{{p_1}{h_2} + 1}}.
\end{equation}
\end{spacing}
\noindent The corresponding decoding error probability is given by ${\varepsilon _2} = Q\left( {f\left( {{{\gamma_2}},{M},D} \right)} \right)$.

Now, we can formulate the optimization problem under NOMA transmission as:
\vspace{-0.5cm}
\begin{spacing}{0.9}
\begin{subequations}\label{initial-pro2}
\begin{align}
\vspace{-0.5cm}
\mathop {\min }\limits_{\left\{ {{p_1},{p_2}} \right\}} \;\;\;&  {{ \varepsilon }_2}\\
{\rm{s.t.}}\;\;\;& {{\bar \varepsilon }_1} \le \varepsilon _1^{\max },\label{rgegVFDue}\\
&M{p_1} + M{p_2} \le {E_{{\rm{tot}}}},\label{fgrSDSEihre}\\
&p_1\le p_2,\label{jiui}
\end{align}
\end{subequations}
\end{spacing}
\noindent where (\ref{jiui}) represents that more power should be allocated to the user with weaker channel gains.

Similar to the proof of Lemma 1, we can show that the energy constraint in (\ref{fgrSDSEihre}) holds with equality at the optimum point. Then, we study the feasible range of the power allocation $p_1$ to facilitate the search algorithm. The expression of ${{\bar \varepsilon }_1}$ can be reexpressed as
\vspace{-0.2cm}
\begin{equation}\label{wdedwe}
\vspace{-0.1cm}
  {{\bar \varepsilon }_1} = {\varepsilon _1} + ({{\hat\varepsilon_1}} - {\varepsilon _1})\varepsilon _2^1\ge {\varepsilon _1}.
\end{equation}
By using constraints (\ref{rgegVFDue}) and (\ref{wdedwe}), we have ${\varepsilon _1} \le \varepsilon _1^{\max }$. By denoting ${\bar{f}}(\gamma)=f(\gamma,M,D)$, the lower bound of $p_1$ can be derived as
\vspace{-0.2cm}
\begin{spacing}{1.1}
\begin{equation}\label{dewde}
 {p_1} \ge \frac{{{{\bar{f}}^{ - 1}}\left( {{Q^{ - 1}}(\varepsilon _1^{\max })} \right)}}{{{h_1}}} \buildrel \Delta \over = p_1^{{\rm{lb}}}.
\end{equation}
\end{spacing}

From constraint (\ref{jiui}), we know that ${p_1} \le \frac{{{E_{{\rm{tot}}}}}}{{2M}}$. To guarantee the meaningfulness of $\varepsilon _2^1$, the inequality ${\log _2}\left( {1 + \gamma_2^1} \right) \ge D/M$ should hold. Then, we have
\vspace{-0.2cm}
\begin{spacing}{1.0}
\begin{equation}\label{desfearfre}
 {p_1} \le \frac{{{E_{{\rm{tot}}}}{2^{ - \frac{D}{M}}}}}{M} - \frac{1}{{{h_1}}} + \frac{{{2^{ - \frac{D}{M}}}}}{{{h_1}}}.
\end{equation}
\end{spacing}
In addition, to guarantee the meaningfulness of $\varepsilon _2$, the inequality ${\log _2}\left( {1 + \gamma_2} \right) \ge D/M$ should hold, which yields
\vspace{-0.5cm}
\begin{spacing}{1.1}
\begin{equation}\label{desfrfre}
 {p_1} \le \frac{{{E_{{\rm{tot}}}}{2^{ - \frac{D}{M}}}}}{M} - \frac{1}{{{h_2}}} + \frac{{{2^{ - \frac{D}{M}}}}}{{{h_2}}}.
\end{equation}
\end{spacing}
Since $h_1>h_2$, the upper bound of $p_1$ is given by
\vspace{-0.2cm}
\begin{spacing}{1.1}
\begin{equation}\label{dehfhur}
  {p_1} \le \min \left\{ {\frac{{{E_{{\rm{tot}}}}{2^{ - \frac{D}{M}}}}}{M} - \frac{1}{{{h_2}}} + \frac{{{2^{ - \frac{D}{M}}}}}{{{h_2}}},\frac{{{E_{{\rm{tot}}}}}}{{2M}}} \right\}  \buildrel \Delta \over =  p_1^{{\rm{ub}}}.
\end{equation}
\end{spacing}

To further reduce the search complexity, in the following theorem, we prove that constraint (\ref{rgegVFDue}) holds with equality at the optimum point.

\emph{\textbf{Theorem 2}}:  Constraint (\ref{rgegVFDue}) holds with equality at the optimum solution.

\upshape Proof: Please see Appendix \ref{theorem2}. \hfill\rule{2.7mm}{2.7mm}

Based on Theorem 2, we can readily know that the one-dimensional line search algorithm can be used to find the optimal $p_1^\star$.

\vspace{-0.4cm}
\subsection{Relay-assisted transmission}
\vspace{-0.15cm}

In this scheme, the robot acts as a relay that assists the transmission for actuator, where decode-and-forward (DF) relay is assumed at the robot. The packet ID is inserted in the packet head for each device to differentiate their corresponding data information. The whole blocklength is divided into two phases, the broadcast phase with blocklength $m_1$ and the relay phase with blocklength $m_2$, which satisfy the constraint of $m_1+m_2\le M$.

In the first phase, the controller broadcasts a large packet that is a combination of two packets to both devices, where the combined packet size is $2D$. The received signals at both devices are given by
\vspace{-0.75cm}
\begin{spacing}{1.2}
\begin{equation}\label{dhedewawfru}
\begin{array}{l}
{y_{1,1}} = \sqrt {{p_s}} {{\tilde h}_1}{\tilde x_1}  + {n_1},\\
{y_{1,2}} = \sqrt {{p_s}} {{\tilde h}_2}{\tilde x_1} + {n_2},
\end{array}
\end{equation}
\end{spacing}
\noindent where $p_s$ denotes the power allocated to the combined packet, $\tilde x_1$ carries the data information of the combined packet with coding rate $2D/m_1$. Then, the SNR of the robot to decode the combined packet is given by ${\gamma _1} = {p_s}{h_1}$, and the decoding error probability at the robot is given by ${\varepsilon _1} = Q\left( {f\left( {{{\gamma_1}},{m_1},2D} \right)} \right)$.

Since the robot acts as a relay based on the DF mode, if the robot successfully decodes the combined packet, it will forward the actuator's packet to the actuator with coding rate $D/m_2$ in the second phase, and the received signal at the actuator is given by
\vspace{-0.2cm}
\begin{spacing}{1.1}
\begin{equation}\label{dehfhiuh}
\vspace{-0.1cm}
  {y_{2,2}} = \sqrt {{p_r}} {{\tilde h}_3}{x_2} + {n_3},
\end{equation}
\end{spacing}
\noindent where $p_r$ is the transmit power at the actuator. The received SNR is ${\gamma _2} ={p_r}{h_3}$, where $h_3$ is the normalized channel gain given by $h_3={{{{\left| {{{\tilde h}_3}} \right|}^2}} \mathord{\left/
 {\vphantom {{{{\left| {{{\tilde h}_3}} \right|}^2}} {\sigma _2^2}}} \right.
 \kern-\nulldelimiterspace} {\sigma _2^2}}$. The  error probability is given by ${\varepsilon _2} = Q\left( {f\left( {{{\gamma_2}},{m_2},D} \right)} \right)$.

There is a possibility that the actuator cannot decode its packet due to the following two reasons: 1) the robot is not able to correctly decode the combined packet and will not forward anything to the actuator with   probability $\varepsilon_1$; and 2) the robot correctly decodes the combined packet and forwards the packet to the actuator with probability $1-\varepsilon_1$, but with probability ${\varepsilon _2}$, the actuator fails to decode the packet. In this case the actuator will have to decode the combined packet by using the received signal from the first phase, i.e., ${y_{1,2}}$.  The achieved SNR of the actuator for decoding the combined packet is given by ${{\hat\gamma_2}} = {{p_s}{h_2}}$, and the corresponding decoding error probability is given by ${{\hat\varepsilon_2}}=Q\left( {f\left( {{{\hat \gamma_2}},{m_1},2D} \right)} \right) $.

As a result, the expected error probability of the actuator decoding its packet in the relay-assisted transmission scheme is given by
\vspace{-0.2cm}
\begin{equation}\label{dewhfurhet}
  {{\bar \varepsilon }_2} = \left( {\left( {1 - \varepsilon _1} \right){\varepsilon _2} + \varepsilon _1} \right)\hat\varepsilon_2.\vspace{-0.4cm}
\end{equation}
Then, the resource allocation problem is formulated as
\vspace{-0.6cm}
\begin{spacing}{0.9}
\begin{subequations}\label{inidewdwpro}
\begin{align}
\vspace{-0.5cm}
\mathop {\min }\limits_{\left\{ {{m_1},{m_2},{p_s},{p_r}} \right\}} \;\;\;&  {{\bar  \varepsilon }_2}\\
{\rm{s.t.}}\;\;\;& {{ \varepsilon }_1} \le \varepsilon _1^{\max },\label{rgdefgue}\\
&{m_1}{p_s} + {m_2}{p_r} \le {E_{{\rm{tot}}}},\label{fgrtrefihre}\\
&m_1+m_2\le M,\label{freerfersdi}\\
& m_1, m_2\in\mathbb{Z}.\label{ofrreffrerpo}
\end{align}
\end{subequations}
\end{spacing}

By using the contradiction method, we can easily prove that constraint (\ref{fgrtrefihre}) holds with equality at the optimal solution. However, in contrast to the above two transmission schemes, the decoding error probability constraint (\ref{rgdefgue}) may not hold with equality at the optimal solution, as the objective function may also decrease with ${ \varepsilon }_1$. The algorithms proposed for the OMA and NOMA transmission schemes cannot be applied.

By using the similar iterative procedure in OMA scheme, we are able to obtain the feasible region of $m_1$ as  $m_1^{\rm{lb}}\le m_1\le m_1^{\rm{ub}}$. For given $m_1$, the search region of $m_2$ can also be obtained as $m_2^{\rm{lb}}(m_1)\le m_2\le (M-m_1)$.

In the following, we study the optimization problem of the power allocation $p_s$ and $p_r$ under fixed $m_1$ and $m_2$. For each given $m_2$, we can obtain the lower bound of $p_r$ to make $\varepsilon_2$ meaningful: ${p_r} \ge {{\left( {{2^{{D \mathord{\left/
 {\vphantom {D {{m_2}}}} \right.
 \kern-\nulldelimiterspace} {{m_2}}}}} - 1} \right)} \mathord{\left/
 {\vphantom {{\left( {{2^{{D \mathord{\left/
 {\vphantom {D {{m_2}}}} \right.
 \kern-\nulldelimiterspace} {{m_2}}}}} - 1} \right)} {{h_3}}}} \right.
 \kern-\nulldelimiterspace} {{h_3}}}\buildrel \Delta \over = p_r^{{\rm{lb}}}$. Thus, the upper bound of $p_s$ can be derived as
\vspace{-0.2cm}
\begin{equation}\label{sdewfr}
{p_s} \le \frac{{{E_{{\rm{tot}}}}}}{{{m_1}}} - \frac{{{m_2}}}{{{m_1}}}p_r^{{\rm{lb}}} \buildrel \Delta \over = p_s^{{\rm{up}}}.
\vspace{-0.2cm}
\end{equation}
Hence, the feasible region of ${p_s}$ is given by $p_s^{\rm{lb}}\le p_s\le p_s^{\rm{ub}}$, where $p_s^{\rm{lb}}$ is the solution to equation ${\varepsilon _1}({p_s}) = \varepsilon _1^{\max }$ with given $m_1$. When $p_s$ is given, $p_r$ can be calculated as $p_r={{\left( {{E_{{\rm{tot}}}} - {m_1}{p_s}} \right)} \mathord{\left/
 {\vphantom {{\left( {{E_{{\rm{tot}}}} - {m_1}{p_s}} \right)} {{m_2}}}} \right.
 \kern-\nulldelimiterspace} {{m_2}}}$. Then, the original optimization problem reduces to a one-dimension  optimization problem  as
\vspace{-0.2cm}
\begin{subequations}\label{idefrefwpro}
\begin{align}
\mathop {\min }\limits_{{p_s} } \;\;\;&  {{\bar  \varepsilon }_2}\\
{\rm{s.t.}}\;\;\;& p_s^{\rm{lb}}\le p_s\le p_s^{\rm{ub}}.\label{rgdgrtdggue}
\end{align}
\end{subequations}
The one-dimensional line search method can be used to solve Problem (\ref{idefrefwpro}).

In summary, we provide Algorithm \ref{hgjoijojoj} to solve Problem (\ref{inidewdwpro}).

\vspace{-0.2cm}
\begin{algorithm}[h]
  \setstretch{0.4}
  \caption{Algorithm for Problem (\ref{inidewdwpro}) }\label{hgjoijojoj}
  \SetKwInOut{Input}{Input}\SetKwInOut{Output}{Output}
 \Input  { $h_1, h_2, D, M, \varepsilon _1^{\max }, E_{{\rm{tot}}}$}
  \Output  {$p_s^\star, p_r^\star, m_1^\star, m_2^\star$}
  Apply the iterative procedure to calculate $m_1^{\rm{lb}}, m_1^{\rm{ub}}$ and $m_2^{\rm{lb}}$;

  \For { $m_1=m_1^{\rm{lb}}: m_1^{\rm{ub}}$ }
  {Calculate the solution to the equation ${{ \varepsilon }_1} =\varepsilon _1^{\max }$, which is denoted as $p_s^{\rm{lb}}$. Calculate the lower bound of $m_2$ with given $m_1$, denoted as $m_2^{\rm{lb}}(m_1)$.

    \For {$m_2=\tilde m_2^{\rm{lb}}:( M-m_1)$}
  {Calculate the upper bound of $p_s$ as $p_s^{\rm{ub}}$ in (\ref{sdewfr}), and solve Problem (\ref{idefrefwpro}). Calculate the objective value ${{\bar \varepsilon }_2}(m_1,m_2)$.
  }
   Given $m_1$, find the blocklength $m_2$ with the minimum value of ${ \varepsilon }_2(m_1, m_2)$:
  \[{\left. {m_2^\#} \right|_{{m_1}}} = \mathop {\arg \min }\limits_{\tilde m_2^{{\rm{lb}}} \le {m_2} \le M - {m_1}} {\varepsilon _2}\left( {{m_1},{m_2}} \right).\]
  }Return $m_1^ \star  = \mathop {\arg \min }\limits_{m_1^{{\rm{lb}}} \le {m_1} \le m_1^{{\rm{ub}}}} {\varepsilon _2}\left( {{m_1},{{\left. {m_2^\#} \right|}_{{m_1}}}} \right),m_2^ \star  = {\left. {m_2^\#} \right|_{m_1^ \star }}$ and the corresponding $p_s^\star$ and $p_r^\star$.
\end{algorithm}
\vspace{-0.2cm}

\subsection{C-NOMA transmission}
\vspace{-0.1cm}

In this part, we consider the C-NOMA transmission in \cite{yuanwei2016}, which is a combination of the NOMA scheme and relay-assisted scheme.  Specifically, in the first phase, the controller transmits two signals $x_1$ and $x_2$ to the two devices via the NOMA technique. In the second phase, the robot acts as a relay and forwards the packet to the actuator. The blocklength for these two phases are denoted by $m_1$ and $m_2$, which satisfies $m_1+m_2\le M$.

Specifically, in the first phase, the received signals at the robot and the actuator are given by
\vspace{-0.3cm}
\begin{spacing}{1.2}
\begin{equation}\label{dhewufru}
\begin{array}{l}
{y_{1,1}} = \sqrt {{p_1}} {{\tilde h}_1}{x_1} + \sqrt {{p_2}} {{\tilde h}_1}{x_2} + {n_1},\\
{y_{1,2}} = \sqrt {{p_1}} {{\tilde h}_2}{x_1} + \sqrt {{p_2}} {{\tilde h}_2}{x_2} + {n_2},
\end{array}
\end{equation}
\end{spacing}
\noindent respectively, where $p_1$ and $p_2$ are the transmit power allocated to the robot and the actuator, $x_1$ and $x_2$ carries different information knowledge for different packets with size $D$. Hence, the coding rate for the transmission to the robot and the actuator are given by $D/m_1$.

By using the NOMA scheme,  the SIC technique is employed at the robot side to cancel the interference from the actuator. Similar to the analysis in the NOMA  scheme, the decoding error probability of $x_2$ at the robot is given by
\vspace{-0.3cm}
\begin{spacing}{1.1}
\begin{equation}\label{wdkrfswedewfok}
\vspace{-0.1cm}
  {{\varepsilon _2^1}}= Q\left( {f\left( {{{\gamma_2^1}},{m_1},D} \right)} \right),
\end{equation}
\end{spacing}
\noindent where $\gamma _2^1$ is the same as that in (\ref{denbbvf}).  Under perfect SIC condition, the decoding error probability of $x_1$ at the robot is given by
\vspace{-0.6cm}
\begin{spacing}{1.1}
\begin{equation}\label{dwdewdjfiehiugh}
{\varepsilon _1}=Q\left( {f\left( {{{\gamma_1}},{m_1},D} \right)} \right),
\end{equation}
\end{spacing}
\noindent where ${\gamma _1} = {p_1}{h_1}$. However, if SIC fails, the corresponding decoding error probability of $x_1$ at the robot is given by   ${\hat\varepsilon_1}=Q\left( {f\left( {{{\hat\gamma_1}},{m_1},D} \right)} \right)$, where $\hat\gamma_1$ is given by (\ref{dertthr}). Using the same analysis as in NOMA, the average decoding probability at the robot is given by
\vspace{-0.2cm}
\begin{spacing}{1.1}
\begin{equation}\label{dewjoihrf}
\vspace{-0.1cm}
  {{\bar \varepsilon }_1} = {\varepsilon _1}(1 - \varepsilon _2^1) + {{{\hat\varepsilon_1}}}\varepsilon _2^1.
\end{equation}
\end{spacing}

By using the similar analysis as in the relay-assisted scheme, the decoding error probability of the actuator decoding $x_2$ under the C-NOMA scheme is given by
\vspace{-0.22cm}
\begin{equation}\label{dewhfurhetjfriehfoi}
\vspace{-0.1cm}
  {{\bar \varepsilon }_2} = \left( {\left( {1 - \varepsilon _2^1} \right){\varepsilon _2} + \varepsilon _2^1} \right)\hat\varepsilon_2,
\end{equation}
where $\varepsilon _2^1$ and $\varepsilon _2$ are given in Subsection-\ref{NOAMscheme}, and $\hat\varepsilon_2$ is the decoding error probability of the actuator when the actuator has to decode $x_2$ from the received signal in the first phase. The expression of  $\hat\varepsilon_2$ is given by
\vspace{-0.2cm}
\begin{equation}\label{wdkoeok}
  {\hat\varepsilon_2}=Q\left(   {f\left( {{\hat\gamma_2},{m_1},D} \right)}\right),
\end{equation}
where ${{\hat\gamma_2}}$ is given by
\vspace{-0.5cm}
\begin{spacing}{1.2}
\begin{equation}\label{edwaew}
  {{\hat\gamma_2}} = \frac{{{p_2}{h_2}}}{{{p_1}{h_2} + 1}}.
\end{equation}
\end{spacing}
Therefore, the optimization problem of C-NOMA transmission scheme can be formulated as
\vspace{-0.5cm}
\begin{spacing}{0.9}
\begin{subequations}\label{initial-cpro2}
\begin{align}
\vspace{-0.2cm}
\mathop {\min }\limits_{\left\{ {{m_1},{m_2},{p_1},{p_2},{p_r}} \right\}} \;\;\;&  {{\bar \varepsilon }_2}\\
{\rm{s.t.}}\;\;\;& {{\bar \varepsilon }_1} \le \varepsilon _1^{\max },\label{rhfiue}\\
& {m_1}({p_1} + {p_2}) + {m_2}{p_r} \le {E_{{\rm{tot}}}},\label{fhrihre}\\
& m_1+m_2\le M,\label{freoi}\\
& m_1, m_2\in\mathbb{Z}, \label{oijypo}\\
& p_1 \le p_2. \label{p1p2coop}
\end{align}
\end{subequations}
\end{spacing}

Following the similar proof as Lemma 1, we can show that constraints (\ref{rhfiue}) and (\ref{fhrihre}) hold with equality at the optimal point, thus the search method can be used to find the optimal solution of Problem (\ref{initial-cpro2}). To reduce the search complexity, we need to find tight lower and upper bounds on $m_1$ and $m_2$.

 However, unlike the previous schemes that only two power allocation variables are involved, the number of power allocation variables in C-NOMA scheme is three. This will complicate the analysis of deriving the bounds of $m_1$ and $m_2$. To deal with this difficulty, we regard the summation of $p_1+p_2$ as a whole entity. {To realize the functionality of the C-NOMA scheme, $\varepsilon_1$ and ${{\varepsilon _2^1}}$ should be very small, e.g., much lower than 0.5. } Then, we have
\vspace{-0.2cm}
\begin{eqnarray}
\vspace{-0.8cm}
{p_1} &\ge& \frac{1}{{{h_1}}}\left( {{2^{\frac{D}{{{m_1}}}}} - 1} \right),\label{jtefrergt}\\
{p_2} &\ge& \frac{1}{{{h_1}}}\left( {{2^{\frac{D}{{{m_1}}}}} - 1} \right)\left( {1 + {p_1}{h_1}} \right).\label{ffearfegt}
\vspace{-0.2cm}
\end{eqnarray}
{By substituting (\ref{jtefrergt}) into the right hand side of (\ref{ffearfegt}), we have
\begin{equation}\label{frgtyju}
  {p_2} \ge \frac{1}{{{h_1}}}\left( {{2^{\frac{D}{{{m_1}}}}} - 1} \right)2^{\frac{D}{{{m_1}}}}.
\end{equation}
By adding (\ref{jtefrergt}) and (\ref{frgtyju}), one can obtain
\vspace{-0.2cm}
\begin{equation}\label{nmjkioijjy}
\vspace{-0.1cm}
  {p_1} + {p_2} \ge \frac{1}{{{h_1}}}\left( {{2^{\frac{{2D}}{{{m_1}}}}} - 1} \right).
\end{equation}}
To ensure that $\varepsilon_2$ is meaningful, we have
\begin{equation}\label{edFGEGT}
  {p_r}  \ge  \frac{1}{{{h_3}}}\left( {{2^{\frac{D}{{{m_2}}}}} - 1} \right).
\end{equation}

 By using the similar iterative procedure, we can also obtain the lower bounds of $m_1$ and $m_2$, which are denoted as $m_1^{{\rm{lb}}}$ and $m_2^{{\rm{lb}}}$, respectively. As a result, the search region of $m_1$ is given by $m_1^{{\rm{lb}}}\le m_1\le (M-m_2^{{\rm{lb}}}) \buildrel \Delta \over = m_1^{{\rm{ub}}}$. For each given $m_1$ within the range, we need to find the search range of $m_2$, which is detailed as follows. Since ${\varepsilon _1} < {{\bar \varepsilon }_1} \le \varepsilon _1^{\max }$, the lower bound of $p_1$ can be obtained by solving the equation of ${\varepsilon _1}({p_1}) = \varepsilon _1^{\max }$ for given $m_1$, which is denoted as $p_1^{\rm{lb}}$. By using (\ref{ffearfegt}), we can obtain the lower bound of $p_2$ as follows:
\vspace{-0.2cm}
\begin{spacing}{1.2}
\begin{equation}\label{ewdewfrf}
\vspace{-0.1cm}
 {p_2} \ge \frac{1}{{{h_1}}}\left( {{2^{\frac{D}{{{m_1}}}}} - 1} \right)\left( {1 + {p_1^{\rm{lb}}}{h_1}} \right) \buildrel \Delta \over = p_2^{\rm{lb}}.
\end{equation}
\end{spacing}
Based on (\ref{fhrihre}), we have
\vspace{-0.35cm}
\begin{spacing}{1.1}
\begin{equation}\label{dewref}
\vspace{-0.05cm}
 {E_{{\rm{tot}}}} - {m_1}(p_1^{{\rm{lb}}} + p_2^{{\rm{lb}}}) \ge {m_2}{p_r} \ge \frac{{{m_2}}}{{{h_3}}}\left( {{2^{\frac{D}{{{m_2}}}}} - 1} \right),
\end{equation}
\end{spacing}
\noindent
where the last inequality is due to the fact that ${p_r} \ge \frac{1}{{{h_3}}}\left( {{2^{\frac{D}{{{m_2}}}}} - 1} \right)$ must hold to guarantee the meaningfulness of ${\varepsilon _2}$. The lower bound of $m_2$ under given $m_1$ (denoted as $m_2^{{\rm{lb}}}(m_1)$) can be obtained from (\ref{dewref}), which is the minimum integer that satisfies  (\ref{dewref}). Obviously, the upper bound of $m_2$ with given $m_1$ is $M-m_1$. Hence, the search region of $m_2$ is given by $m_2^{{\rm{lb}}}(m_1)\le m_2\le (M-m_1)$.

Given $m_1$ and $m_2$, we need to find the optimal $p_1$, $p_2$ and $p_s$. These variables are coupled and it is difficult to find  the optimal solution by using the  optimization  method. The one-dimensional search is adopted to find the optimal solution. In particular, we first fix the value of the sum of $p_1$ and $p_2$ as $t$, i.e., $t=p_1+p_2$. Since constraint (\ref{rhfiue}) holds with equality at the optimal point, the optimal $p_1$ can be obtained by solving
the equation $\bar \varepsilon_1(p_1)=\varepsilon_1^{\rm{max}}$ by inserting $p_2=t-p_1$ into this equation. By combining (\ref{ffearfegt}) and (\ref{p1p2coop}), the upper bound of $p_1$ is obtained as $p_1 \le \min{\left( t \cdot 2^{-\frac{D}{m_1}}-\frac{1}{h_1}+\frac{1}{h_1} \cdot 2^{-\frac{D}{m_1}},\frac{t}{2}\right)}\triangleq p_1^{\rm{up}}$, and $p_1$ should be within the domain $p_1 \in (p_1^{\rm{lb}},p_1^{\rm{up}})$. This equation has only one variable $p_1$ and the one-dimensional search method can be adopted to solve the equation. As constraint (\ref{fhrihre}) holds with equality, $p_r$ can be directly obtained as $p_r={{({E_{{\rm{tot}}}} - t{m_1})} \mathord{\left/
 {\vphantom {{({E_{{\rm{tot}}}} - t{m_1})} {{m_2}}}} \right.
 \kern-\nulldelimiterspace} {{m_2}}}$.  Calculate the objective value with given $m_1$, $m_2$, $t$ and $p_r$.  The remaining task is to find the tight search region $t$. Obviously, the lower bound of $t$ is given by $t^{\rm{lb}}=p_1^{{\rm{lb}}}+p_2^{{\rm{lb}}}$. To obtain the upper bound of $t$, we first find the lower bound of $m_2p_r$, which is  given by
 \vspace{-0.1cm}
 \begin{spacing}{1.0}
 \begin{equation}\label{rfegfwtg}
 \vspace{-0.1cm}
 {m_2}{p_r} \ge \frac{{{m_2}}}{{{h_3}}}\left( {{2^{\frac{D}{{{m_2}}}}} - 1} \right).
 \end{equation}
 \end{spacing}
 Then, the upper bound of $t$ is given by
 \vspace{-0.2cm}
 \begin{spacing}{1.0}
 \begin{equation}\label{dewafre}
   t \le \frac{1}{{{m_1}}}\left( {{E_{{\rm{tot}}}} - \frac{{{m_2}}}{{{h_3}}}\left( {{2^{\frac{D}{{{m_2}}}}} - 1} \right)} \right) = {t^{{\rm{ub}}}}.
 \end{equation}
 \end{spacing}
Based on the above analysis, we provide Algorithm \ref{hgjoj} to solve Problem (\ref{inidewdwpro}).

\begin{algorithm}
\setstretch{0.4}
  \caption{Algorithm for Problem (\ref{initial-cpro2}) }\label{hgjoj}
  \SetKwInOut{Input}{Input}\SetKwInOut{Output}{Output}
 \Input  { $h_1, h_2, h_3, D, M, \varepsilon _1^{\max }, E_{{\rm{tot}}}$}
  \Output  {$p_1^\star, p_2^\star, p_r^\star, m_1^\star, m_2^\star$}
  Apply the iterative procedure to calculate $m_1^{\rm{lb}}, m_1^{\rm{ub}}$ and $m_2^{\rm{lb}}$;

  \For { $m_1=m_1^{\rm{lb}}: m_1^{\rm{ub}}$ }
  {Calculate the solution to the equation ${{ \varepsilon }_1} =\varepsilon _1^{\max }$, which is denoted as $p_1^{\rm{lb}}$. Use (\ref{ewdewfrf}) to calculate the lower bound of $p_2$, denoted as $p_2^{\rm{lb}}$. Use
  (\ref{dewref}) to find the lower bound of $m_2$, denoted as $m_2^{{\rm{lb}}}(m_1)$.

  \If {$m_2^{{\rm{lb}}}(m_1)\le (M-m_1)$ }
    {  \For {$m_2=m_2^{{\rm{lb}}}(m_1):( M-m_1)$}
  {Calculate the lower bound of $t$ as $t^{\rm{lb}}=p_1^{{\rm{lb}}}+p_2^{{\rm{lb}}}$ , and the upper bound of $t$ as $t^{\rm{ub}}$ from (\ref{dewafre}).  Use the one-dimensional search to find the optimal $t$ that achieves the minimum objective value. Denote the optimal objective value ${{\bar \varepsilon }_2}(m_1,m_2)$.
  }
   Given $m_1$, find the blocklength $m_2$ with the minimum value of ${ \varepsilon }_2(m_1, m_2)$:
  \[{\left. {m_2^\#} \right|_{{m_1}}} = \mathop {\arg \min }\limits_{m_2^{{\rm{lb}}}(m_1) \le {m_2} \le M - {m_1}} {\varepsilon _2}\left( {{m_1},{m_2}} \right).\]
  }
  }
  Return $m_1^ \star  = \mathop {\arg \min }\limits_{m_1^{{\rm{lb}}} \le {m_1} \le m_1^{{\rm{ub}}}} {\varepsilon _2}\left( {{m_1},{{\left. {m_2^\#} \right|}_{{m_1}}}} \right),m_2^ \star  = {\left. {m_2^\#} \right|_{m_1^ \star }}$ and the corresponding $p_1^\star$ and $p_2^\star$.
\end{algorithm}

  \textbf{Remark:} It is noted that the feasible region of C-NOMA scheme is smaller than that of the relay-assisted transmission scheme. Specifically, if $p_1^*$ and $p_2^*$ is any one feasible solution of Problem (\ref{initial-cpro2}), it can be readily checked that $p_s=p_1^*+p_2^*$ is also a feasible solution of Problem (\ref{inidewdwpro}). However, if $\{p_1^*,p_2^*\}$  is not a feasible solution of Problem (\ref{initial-cpro2}), $p_s=p_1^*+p_2^*$ may still be feasible for Problem (\ref{inidewdwpro}). For example, by letting
\vspace{-0.15cm}
\begin{spacing}{1.0}
\begin{equation}\label{dwefrswq}
 {p_2} = \frac{1}{{{h_2}}}\left( {{2^{\frac{D}{{{m_1}}}}} - 1} \right),{p_1} = \frac{1}{{{h_1}}}\left( {{2^{\frac{{2D}}{{{m_1}}}}} - 1} \right) - \frac{1}{{{h_2}}}\left( {{2^{\frac{D}{{{m_1}}}}} - 1} \right),
\end{equation}
\end{spacing}
\noindent it can be readily checked that $p_1$ and $p_2$  do not satisfy condition (\ref{ffearfegt}), which is not feasible for Problem (\ref{initial-cpro2}). However by setting $p_s=p_1+p_2$, $p_s$ is still feasible for Problem (\ref{inidewdwpro}). This observation means the feasible region for Problem (\ref{inidewdwpro}) is larger than that of Problem (\ref{initial-cpro2}).

{\section{Extension to More Devices for the OMA Scheme}\label{jfejoref}
In this section, we consider the more general case when the system has more than two devices for the OMA scheme. The extension to other schemes will be studied in the future work.
\subsection{Sytem Model and Problem Formulation}

Let us denote the total number of devices as $K$, and the set of all devices as $\cal K$. We assume that the normalized channel gains  of all $K$ devices are arranged in a decreasing order, i.e.,
$h_1>h_2>\cdots >h_K$\footnote{{Due to the small-scale fading, the probability that any two or more devices have the same channel gain is equal to zero.}}. Then, we aim to jointly optimize the power and blocklength allocation to minimize the decoding error probability of the $K$th device while guaranteeing the decoding error probability requirements of the first $K-1$ devices. Mathematically, the optimization problem can be formulated as follows:
\begin{subequations}\label{Ktarget}
\begin{align}
\min_{\left\{m_k, k\in \cal K\right\},\left\{p_k,k\in \cal K \right\}}\;\;\;\;&\varepsilon_K\\
\text{s.t.}\;\;\;\;\;&\varepsilon_k\le\varepsilon_k^{\max},\;\; k\in {\cal K}\backslash K,\label{C1}\\
&\sum\nolimits_{k\in \cal K}m_kp_k\le E_{\text{tot}},\label{C2}\\
&\sum\nolimits_{k\in \cal K}m_k\le M,\label{C3}\\
& m_k \in\mathbb{Z},k\in \cal K. \label{oxasetypo}
\end{align}
\end{subequations}
In contrast to the case of two devices where the globally optimal solution to Problem (\ref{initial-pro1}) can be obtained,     the globally optimal solution to Problem (\ref{Ktarget}) for the more general case is not available. In the following, we aim to obtain a suboptimal solution to Problem (\ref{Ktarget}).

\subsection{Problem Reformulation}
To make Problem (\ref{Ktarget}) tractable, we again approximate $V$ as one, i.e., $V \approx 1$. This approximation is very accurate when the SNR value $\gamma$ is very high, i.e., $\gamma\gg 1$. As the decoding error probability is a decreasing function of power and blocklength, we can readily prove that constraints  (\ref{C1}), (\ref{C2}) and (\ref{C3}) hold with equality at the optimum point by using the contradiction method. By using the fact that $\varepsilon_k=\varepsilon_k^{\max}, k \in {\cal K}\backslash K$, $p_k$  can be derived as a function of  $m_k$, given by
\begin{equation}\label{acfref}
  p_k=\frac{2^{\frac{D}{m_k}+\frac{Q^{-1}(\varepsilon_k^{\max})}{\ln2\sqrt{m_k}}}-1}{h_k}\triangleq \chi(m_k), {k\in \cal K}\backslash K.
\end{equation}
By substituting (\ref{acfref}) into (\ref{Ktarget}), Problem (\ref{Ktarget}) can be transformed as follows:
\begin{subequations}\label{Ktxscdet}
\begin{align}
\min_{\left\{m_k, k\in \cal K\right\}, p_K  }\;\;\;\;&\varepsilon_K\\
\text{s.t.}\;\;\;\;\;&\sum\limits_{k \in {\cal K}\backslash K}m_k\chi(m_k)+m_Kp_K= E_{\text{tot}},\label{Casxs}\\
&\sum\limits_{k\in {\cal K}}m_k= M, m_k \in\mathbb{Z},k\in \cal K. \label{Caxas}
\end{align}
\end{subequations}
Compared with the original Problem (\ref{Ktarget}), the number of optimization variables of Problem (\ref{Ktxscdet}) is significantly reduced. However, this problem is still difficult to solve. In the following, we first use the  exhaustive search to find $m_K$, and then  optimize $p_K$. To this end, we need to find tight lower and upper bounds of $m_K$ to reduce the computational complexity.

\subsection{Bounds of $m_K$}

In this subsection, we attempt to obtain the bounds of $m_K$. We first provide the following theorem.

\emph{\textbf{Theorem 3}}: Define $A_k={{{Q^{ - 1}}(\varepsilon _k^{\max })} \mathord{\left/
 {\vphantom {{{Q^{ - 1}}(\varepsilon _k^{\max })} {\ln 2}}} \right.
 \kern-\nulldelimiterspace} {\ln 2}}$ and $g\left( {{m_k}} \right)  \buildrel \Delta \over =  {m_k}\chi ({m_k})$. Then, $g\left( {{m_k}} \right) $ is a monotonically decreasing and convex function when $m_k$ satisfies:
 \begin{equation}\label{fdvtgxdtr}
  \sqrt {{m_k}}  < \frac{{\frac{3}{4}{A_k}\ln 2 + \sqrt {\frac{9}{{16}}{{\left( {\ln 2} \right)}^2}A_k^2 + 8D\ln 2} }}{2}.
 \end{equation}
\emph{Proof}: \upshape Please see Appendix \ref{theorem3}. \hfill\rule{2.7mm}{2.7mm}

In general, for a typical URLLC system, the number of transmission bits is around $100$ bits and the decoding error probability requirement is around $10^{-9}$. Then, $A_k$ is 8.653, and the value of the right hand side of (\ref{fdvtgxdtr}) is given by 14.236. Then, when $m_k\le 202$,  the inequality (\ref{fdvtgxdtr}) holds. In short packet transmission with OMA scheme, the number of blocklength to each device is generally smaller than 100. Hence, in our considered scenario,   $g\left( {{m_k}} \right)$ can be regarded as a monotonically decreasing and convex function.

In the following, we provide an iterative procedure to obtain the tight bounds of $m_K$. Since $m_kp_k=g(m_k)<E_{\rm{tot}}$ and $g(m_k)$ is a monotonically decreasing function, we can obtain the lower bound of $m_k$ by using the bisection search method, which is denoted as $m_k^{\rm{lb}(0)},k\in {\cal K}\backslash K$. To guarantee the meaningfulness of ${\varepsilon _K}$, the following inequality holds
 \begin{equation}\label{defgtr}
   {p_K} > {{\left( {{2^{\frac{D}{{{m_K}}}}} - 1} \right)} \mathord{\left/
 {\vphantom {{\left( {{2^{\frac{D}{{{m_K}}}}} - 1} \right)} {{h_K}}}} \right.
 \kern-\nulldelimiterspace} {{h_K}}}.
 \end{equation}
Then, we have
\begin{equation}\label{wdREF}
  {E_{{\rm{tot}}}} > {m_K}{p_K} > \frac{{{m_K}}}{{{h_K}}}\left( {{2^{\frac{D}{{{m_K}}}}} - 1} \right)\buildrel \Delta \over = q({m_K}).
\end{equation}
As a result, we can obtain the lower bound of $m_K$ from (\ref{wdREF}), which is denoted as $m_K^{\rm{lb}(0)}$. Then, for each device $k$, the upper bound of $m_k$ is given by $m_k^{\rm{ub}(0)}= M - \sum\nolimits_{i \in {\cal K}\backslash k} {m_i^{{\rm{lb}}(0)}}, k\in \cal K$. Since $q({m_K})$ defined in (\ref{wdREF}) is a monotonically decreasing function, we have $q({m_K})>q({m_K^{\rm{ub}(0)}})$. In addition, $g(m_k)$ is a monotonically decreasing function of $m_k$, and we have $g(m_k)>g(m_k^{\rm{ub}(0)}), k\in {\cal K}\backslash K$. Then, for each $k\in {\cal K}\backslash K$, we have
\begin{equation}\label{xssawsavg}
  {E_{{\rm{tot}}}} - \sum\nolimits_{i \in {\cal K}\backslash \{K,k\}} {g(m_i^{{\rm{ub}(0)}})}-q({m_K^{\rm{ub}(0)}})> g(m_k), k\in {\cal K}\backslash K.
\end{equation}
Then, the lower bound of $m_k$ for $k\in {\cal K}\backslash K$ can be obtained as ${m_k^{{\rm{lb}}(1)}}, k\in {\cal K}\backslash K$.
For the $K$th device, we have
\begin{equation}\label{xsfreavg}
  {E_{{\rm{tot}}}} - \sum\nolimits_{k \in {\cal K}\backslash K} {g(m_k^{{\rm{ub}(0)}})}>\frac{{{m_K}}}{{{h_K}}}\left( {{2^{\frac{D}{{{m_K}}}}} - 1} \right).
\end{equation}
Then, based on (\ref{xsfreavg}) we can update the lower bound of $m_K$ as $m_K^{{\rm{lb}(1)}}$. Then, for each device $k$, the upper bound of $m_k$ is given by $m_k^{\rm{ub}(1)}= M - \sum\nolimits_{i \in {\cal K}\backslash k} {m_i^{{\rm{lb}}(1)}}, k\in \cal K$. Finally, repeat the above procedure until   $m_K^{{\rm{lb}}(n)}=m_K^{{\rm{lb}}(n+1)}$ and $m_K^{{\rm{ub}}(n)}=m_K^{{\rm{ub}}(n+1)}$, where $n$ is the iteration number. Similar to the case of two devices, the above procedure can be proved to be convergent, and denote the final converged upper and lower bounds of $m_K$ as $m_K^{\rm{ub}}$ and   $m_K^{\rm{lb}}$, respectively.

\subsection{Optimization of $p_K$ with Given $m_K$}
Given $m_K$,  $\varepsilon_K$ is a monotonically decreasing function of $p_K$  and $p_K$ is given by
 \begin{equation}\label{defrgt}
  {p_K} = \frac{1}{{{m_K}}}\left( {{E_{{\rm{tot}}}} - \sum\limits_{k \in {\cal K}\backslash K} {{m_k}\chi ({m_k})} } \right),
 \end{equation}
Problem (\ref{Ktxscdet}) can then be equivalently transformed as
\begin{subequations}\label{Ktxsdwt}
\begin{align}
\min_{\left\{m_k, k\in {\cal K}\backslash K\right\} }\;\;\;\;&\sum\limits_{k \in {\cal K}\backslash K}m_k\chi(m_k)\\
\text{s.t.}\;\;\;\;\;&\sum\limits_{k \in {\cal K}\backslash K}m_k =M-m_K,\label{edsawxs}\\
& m_k \in\mathbb{Z},k \in {\cal K}\backslash K. \label{Cadxswas}
\end{align}
\end{subequations}
This problem is still difficult to solve due to  the integer constraint (\ref{Cadxswas}). To resolve this issue, we relax $\{m_k,k\in {\cal K}\backslash K\}$ to continuous values. Then, Problem (\ref{Ktxsdwt}) can be relaxed as follows:
\begin{subequations}\label{Kadwaqwdwt}
\begin{align}
\min_{\left\{m_k, k\in {\cal K}\backslash K\right\} }\;\;\;\;&\sum\limits_{k \in {\cal K}\backslash K}m_k\chi(m_k)\\
\text{s.t.}\;\;\;\;\;& m_k\ge m_k^{\rm{lb}}, k\in {\cal K}\backslash K, (\ref{edsawxs}),
\end{align}
\end{subequations}
where $\{m_k^{\rm{lb}}, k\in {\cal K}\backslash K\}$ are given in the above subsection. Since $m_k\chi(m_k)$ is proved to be convex as shown in Theorem 2, Problem (\ref{Kadwaqwdwt}) is a convex optimization problem, which  can be solved by using the  Lagrangian dual decomposition method \cite{boyd}. We first introduce the Lagrange multiplier $\lambda$ associated with constraint (\ref{edsawxs}), the partial Lagrangian function of Problem (\ref{Kadwaqwdwt}) is given by
\begin{equation}\label{wdwf}
  {\cal L}({\bf{m}},\lambda ) =\sum\limits_{k \in {\cal K}\backslash K}m_k\chi(m_k) + \lambda \left( \sum\limits_{k \in {\cal K}\backslash K}m_k - M-m_K\right),
\end{equation}
where ${\bf{m}}=\{m_k,k\in {\cal K}\backslash K\}$.

In the following, we aim to obtain the optimal $m_k, k\in {\cal K}\backslash K$ for given $\lambda$, which is denoted as $m_k^\star(\lambda), k\in {\cal K}\backslash K$.
As ${\cal L}({\bf{m}},\lambda ) $ is a convex function of $m_k, k\in {\cal K}\backslash K$, the optimal $m_k$ for given $\lambda$ can be obtained in the following. If
\begin{equation}\label{dqwefrt}
 {\left. {\frac{{\partial {\cal L}({\bf{m}},\lambda )}}{{\partial {m_k}}}} \right|_{{m_k} = m_k^{{\rm{lb}}}}} \ge 0,
\end{equation}
the optimal $m_k$ is given by $m_k^\star(\lambda)  = m_k^{{\rm{lb}}}$. Otherwise, $m_k^\star(\lambda)$ is the solution to the following equation:
\begin{equation}\label{WADEFR}
 \frac{{\partial {\cal L}({\bf{m}},\lambda )}}{{\partial {m_k}}} = 0,
\end{equation}
which can be obtained by using the bisection search method.

Upon obtaining the optimal ${m_k^\star}(\lambda),k\in {\cal K}\backslash K$, we can obtain the value of the left hand side of (\ref{edsawxs}), which is defined as function  $F(\lambda )$
\begin{equation}\label{efrshy}
 F(\lambda ) \buildrel \Delta \over =  \sum\limits_{k \in {\cal K}\backslash K}m_k^\star(\lambda).
\end{equation}
By using the similar technique as in Appendix A of \cite{pan2019intelligent}, we can prove that $F(\lambda )$ is a monotonically decreasing function of $\lambda$. Hence, the bisection search method can be adopted to find the solution of $\lambda$ to the equation $ F(\lambda )=M-m_K$ if the original problem is feasible.

Denote the solution obtained by solving the relaxed problem (\ref{Kadwaqwdwt}) as $\left\{{\bar m}_k, k\in {\cal K}\backslash K\right\}$. In general,   $\left\{{\bar m}_k, k\in {\cal K}\backslash K\right\}$ may violent the integer requirement. Hence, we need to convert the continuous $\left\{{\bar m}_k, k\in {\cal K}\backslash K\right\}$ to integer solutions, denoted as $\left\{{  m}_k^\star, k\in {\cal K}\backslash K\right\}$. However, the integer conversion problem is a combinatorial optimization problem, which is NP to solve. In the following, we apply the greedy search method to solve the integer conversion problem. Specifically, we first initialize the integer solution as $m_k^\star=\left\lfloor {{{\bar m}_k}} \right\rfloor, k\in {\cal K}\backslash K$.  Note that $g(m_k)$ is a monotonically decreasing function of $m_k$. Each time we allocate one blocklength to the device with the largest decrement of $g(m_k)$, i.e., ${k^*} = \mathop {\arg } \max\nolimits_{k \in {\cal K}\backslash K}\left\{ {g({m_k}) - g({m_k} + 1)} \right\}$. The rational behind this is that based on (\ref{edsawxs}) more energy can be allocated to the $K$th device, thus decreasing ${\varepsilon _K}$  most. For the $k^*$th device, we set $m_{k^*}^\star=m_{k^*}^\star+1$.  If $p_K^\star$ is smaller than zero, set ${\varepsilon _K^\star}=1$. Repeat the above procedure until  $\sum\nolimits_{k \in {\cal K}\backslash K}m_k^\star=M-m_K$. Then,   the power allocated to the $K$th device can be recalculated as
\begin{equation}\label{frgtftgf}
  p_K = \frac{{{E_{{\rm{tot}}}} - \sum\limits_{k \in {\cal K}\backslash K} {g(m_k^ \star )} }}{{m_K }}.
\end{equation}
Thus, we can calculate ${\varepsilon _K}$ based on current $m_K$ and $p_K^\star$.}

\vspace{-0.2cm}
\section{Simulations Results}
\vspace{-0.1cm}
\label{simulation}
In this section, simulation results are provided to evaluate the performance of the proposed algorithms. For simplicity, we assume that the controller, the robot and the actuator are located on the same line, and the robot is moving from the controller to the actuator, and the robot is served as the relay to help the transmission of the actuator. The distance between the controller and the actuator is set as $500$ m. Let us denote $d_1$, $d_2$ and $d_3$ as the distances from the controller to the robot, the controller to the actuator, and the robot to the actuator, respectively. The system bandwidth is set as $B=1$ MHz. Hence, the downlink transmission delay duration is calculated as $100\ {\rm{us}}$ that meets a criterion of industrial standards\cite{15iccw}. The noise power spectral density is -173 dBm/Hz. The decoding (packet) error probability  requirement for the robot is set as $10^{-9}$. The large-scale path loss model is  $35.3+37.6\log_{10}{\rm{dB}}$ \cite{access2010further}.
The simulation section is divided into two subsections. In the first subsection, we assume that the channel gain is only determined by the path loss in order to obtain the insights of all the schemes.  In the second subsection, we consider the network availability performance \cite{changyangshetcom2018} taking into account  small-scale fading obeying the Rayleigh distribution.

\vspace{-0.6cm}
\subsection{Only Large-scale Fading}
\vspace{-0.1cm}

\begin{figure}[h]
\begin{minipage}[t]{0.5\linewidth}
\vspace{-0.6cm}
\setcaptionwidth{0.95 \textwidth}
\captionsetup{font=small}
\centering
\includegraphics[width=3.1in]{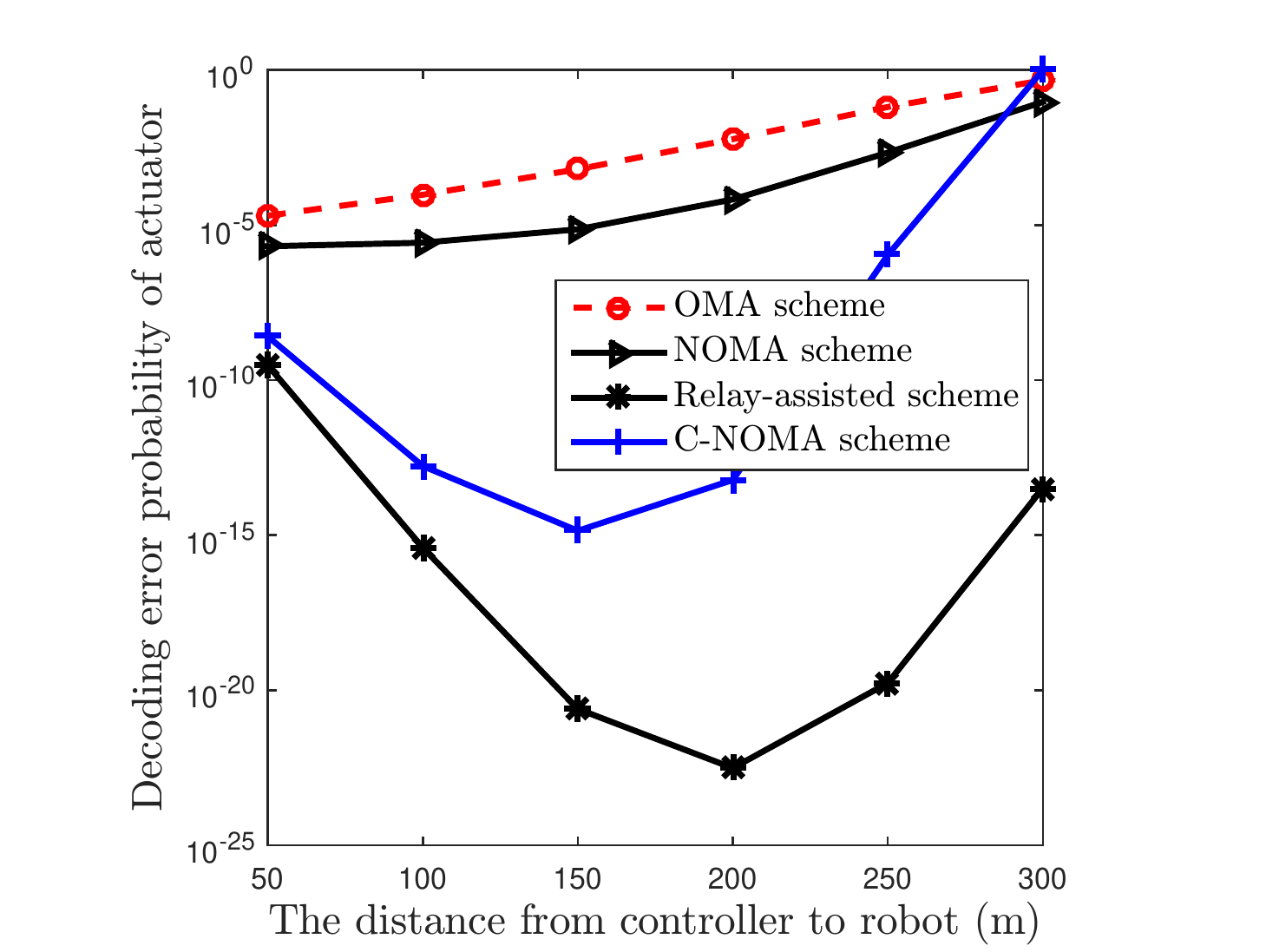}
\vspace{-0.3cm}
\caption{The decoding error probability of the actuator versus the distance from the controller to the robot under four schemes, when $D=100$ bits, $M=100$ symbols, $\tilde{E}_{\rm{tot}}=5\times 10^{-5}$ Joule.}
\vspace{-1cm}
\label{Dis}
\end{minipage}%
\begin{minipage}[t]{0.5\linewidth}
\vspace{-0.6cm}
\captionsetup{font=small}
\setcaptionwidth{0.95 \textwidth}
\centering
%  % Requires \usepackage{graphicx}
%\includegraphics[width=\textwidth]{./fig/ErrvsM.pdf}
\includegraphics[width=3.1in]{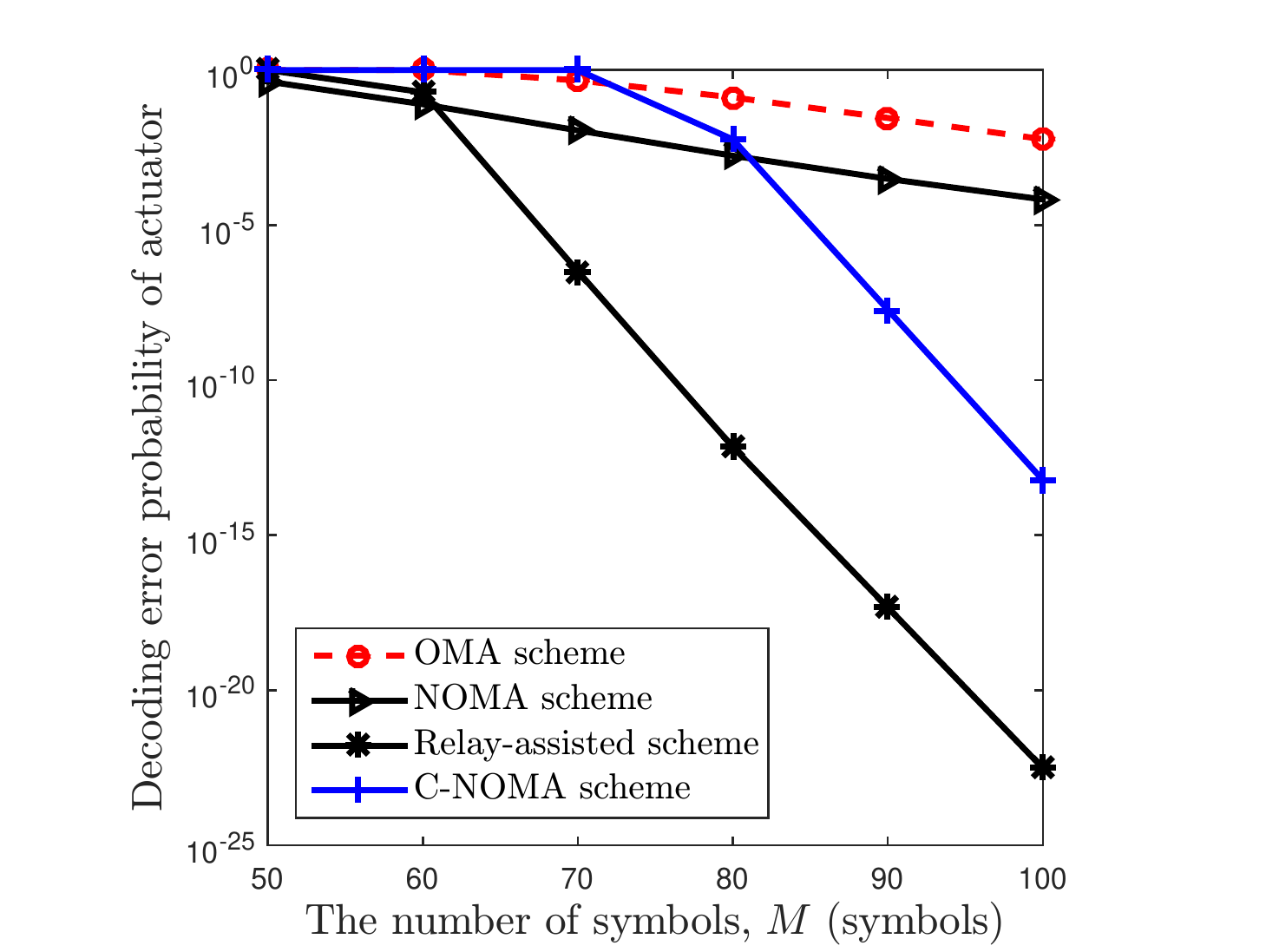}
\vspace{-0.3cm}
\caption{The decoding error probability of the actuator versus the number of symbols under four schemes, when $D=100$ bits, $\tilde{E}_{\rm{tot}}=5\times 10^{-5}$ Joule, $d_1=200$ m, $d_2=500$ m, and $d_3=300$ m.}
\label{blocklength}
\end{minipage}%
\vspace{-0.75cm}
\end{figure}

In Fig.~\ref{Dis}, we first study the impact of distance $d_1$ on the decoding error probability. We observe that relay-assisted transmission outperforms the other three schemes.  It is interesting to see that when the  robot moves from the controller to the actuator, the decoding error probability achieved by the OMA and NOMA schemes always decreases.  The main reason is that the channel gain from controller to  robot decreases with increasing the distance, so the energy and blocklength required for the robot to guarantee its error probability requirement increases. As a result, the available energy and blocklength for the actuator will decrease.  On the other hand, the reliability performances achieved by the C-NOMA and relay-assisted schemes first increase and then decrease when the robot moves in the line. This can be explained as follows. When the robot moves from $50\ {\rm{m}}$ to $150\ {\rm{m}}$ for the C-NOMA and $200\ {\rm{m}}$ for relay-assisted scheme, the channel gain from the robot to the actuator becomes weak, which is the performance bottleneck that limits the decoding error probability of the actuator. However, when the robot continues to move towards the actuator, the transmission link from the controller to the robot becomes the bottleneck link. Hence, the distance $d_1$ can be optimized to additionally improve the system performance, which can be treated in the future work. It is interesting to observe that the C-NOMA performs worse than the relay-assisted scheme, which is due to the larger feasible region for the latter scheme as explained at the end of Section \ref{scheme}.

In Fig.~\ref{blocklength}, we examine the impact of available blocklength $M$ on the decoding error probability of the actuator. As expected, larger $M$ leads to much better reliability performance in all schemes, and the decoding error probability achieved by the relay scheme decreases from $1$ to $10^{-22}$ with $M$ increasing from 50 to 100.  It is interesting to find that when the blocklength $M$ is equal to 50 and 60, the NOMA scheme has the best reliability performance since the whole transmission blocklength can be used for transmission in NOMA, while the whole blocklength should be divided into two parts for the other schemes. Importantly, this provides insights for the system designer that when the blocklength is very limited as in URLLC, relay may not be a good option since some blocklengths needs to be reserved for the two-stage transmission. However, further increasing $M$, the relay-assisted transmission and the C-NOMA start to perform better than the NOMA scheme, and the performance gain monotonically increases with $M$.  However, the cross-point associated with the relay scheme is much lower than that of the C-NOMA scheme due to the shrinking feasible region associated with the latter scheme. Furthermore, the curves of both schemes have the same slope with different bias.

\begin{figure}[h]
\vspace{-0.7cm}
\begin{minipage}[t]{0.5\linewidth}
\setcaptionwidth{0.95 \textwidth}
\captionsetup{font=small}
\centering
\includegraphics[width=3.1in]{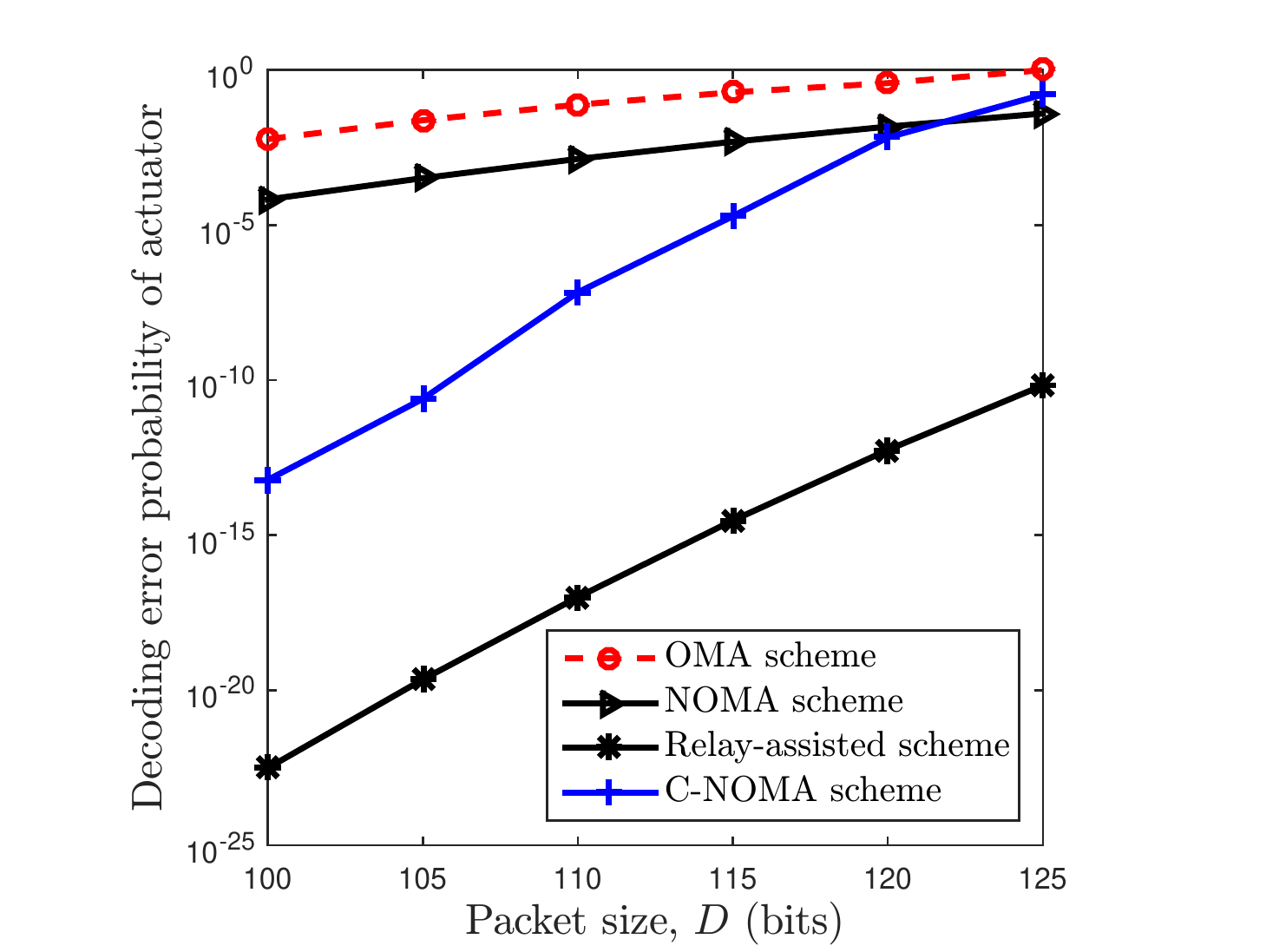}
\vspace{-0.3cm}
\caption{The decoding error probability of the actuator versus packet size under four schemes, when $M=100$ sysmbols, $\tilde{E}_{\rm{tot}}=5\times 10^{-5}$ Joule, $d_1=200$ m, $d_2=500$ m, and $d_3=300$ m.}
\label{packetsize}
\vspace{-1.2cm}
\end{minipage}
\begin{minipage}[t]{0.5\linewidth}
\setcaptionwidth{0.95 \textwidth}
\captionsetup{font=small}
\centering
\includegraphics[width=3.1in]{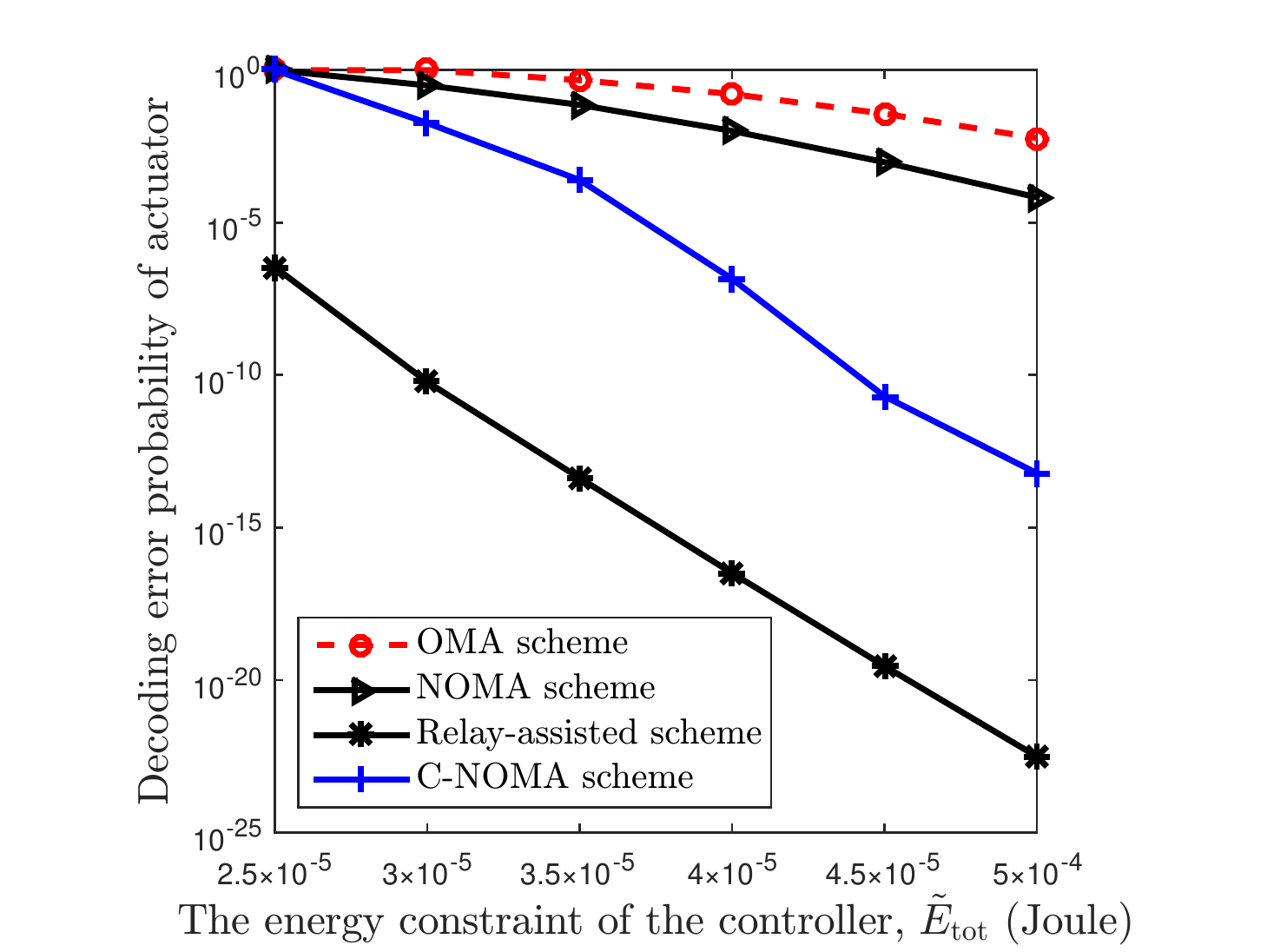}
\vspace{-0.3cm}
\caption{The decoding error probability of the actuator versus the energy constraint under four schemes, when $D=100$ bits, $M=100$ symbols, $d_1=200$ m, $d_2=500$ m, and $d_3=300$ m.}
\vspace{-0.8cm}
\label{Etot}
\end{minipage}
\end{figure}

In Fig.~\ref{packetsize}, we study the impact of the packet size $D$ on the decoding error probability. As expected, a larger packet size leads to a higher error probability for all schemes. The performance advantage of the relay-assisted scheme over the OMA and NOMA schemes shrinks with the increase of $D$. It is interesting to find that the curves associated with the OMA and NOMA schemes have almost the same slope, while those of the relay-assisted transmission and the C-NOMA scheme are similar. The main reason may be that the latter two schemes apply relay to assist the transmission. Similar to the observations in \cite{xiaoyusun}, the NOMA achieves better performance than the OMA scheme. When $D=125$ bits, the C-NOMA is even worse than the NOMA since some blocklengths should be reserved for the two-stage transmission in the former scheme.

In Fig.~\ref{Etot}, we study the impact of the total energy on the decoding error probability. It is observed that more available energy leads to better reliability performance as expected. It is also seen that the relay-assisted transmission has the best performance, and the performance gain increases with the amount of available energy. It is shown that with sufficient energy, transmission with the aid of relay (i.e., the relay-assisted transmission and the C-NOMA transmission) is beneficial for the system performance. When ${\tilde E_{{\rm{tot}}}}= 5\times 10^{-5}$ Joule, the decoding error probability achieved by the relay-assisted transmission is extremely low.

\begin{figure}
\begin{minipage}[t]{0.5\linewidth}
\setcaptionwidth{0.95 \textwidth}
\captionsetup{font=small}
\centering
\includegraphics[width=3.1in]{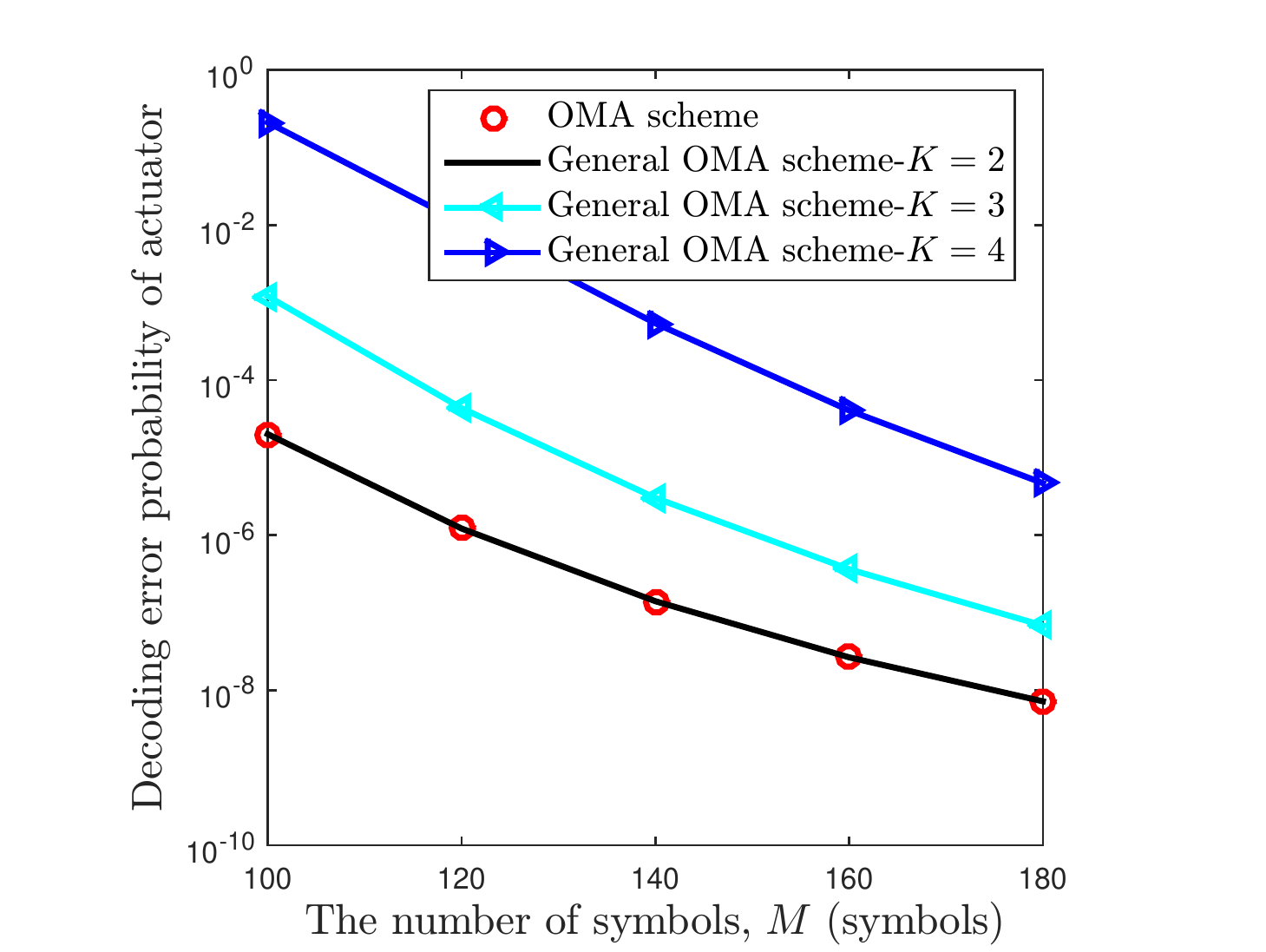}
\vspace{-0.3cm}
\caption{{ The decoding error probability of the actuator versus the number of symbols for the OMA scheme the general OMA scheme, when $D=100$ bits, and $\tilde{E}_{\rm{tot}}=5\times 10^{-5}$ Joule.}}
\label{general}
\vspace{-0.7cm}
\end{minipage}
\begin{minipage}[t]{0.5\linewidth}
\captionsetup{font=small}
\setcaptionwidth{0.95 \textwidth}
\centering
%  % Requires \usepackage{graphicx}
%\includegraphics[width=\textwidth]{./fig/Symbol.pdf}
\includegraphics[width=3.1in]{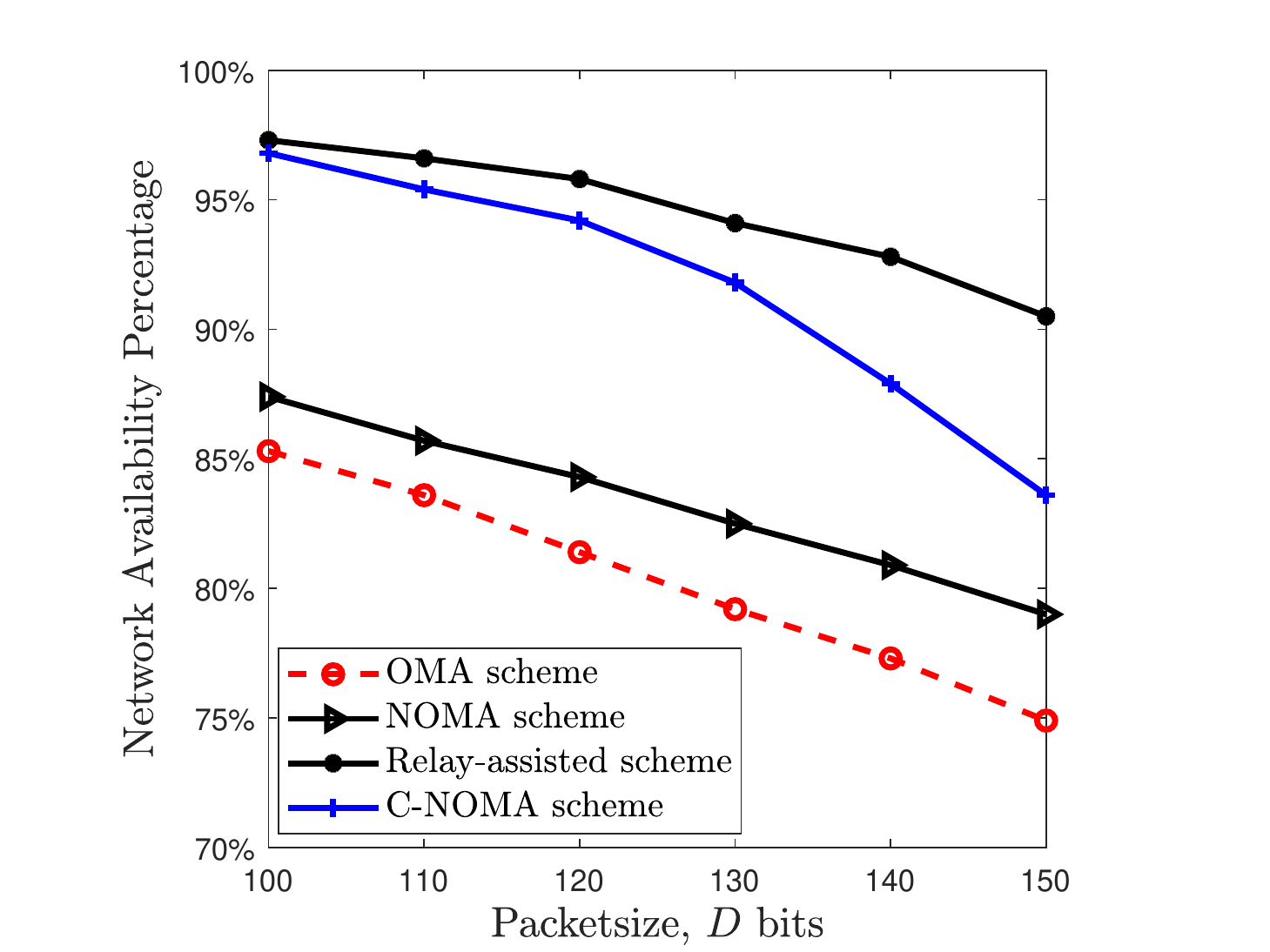}
\vspace{-0.3cm}
\caption{The network availability percentage versus the packet size $D$ under four schemes, when $\tilde{E}_{\rm{tot}}=5\times10^{-4}$ Joule, $M=100$ symbols.}
\label{packetna}
\vspace{-0.5cm}
\end{minipage}%
\vspace{-0.7cm}
\end{figure}

{In Fig.~\ref{general}, we study the performance comparison between the OMA scheme in Section \ref{scheme} and the general OMA  in Section \ref{jfejoref}. Denote the number of devices as $K$. If $K=2$, both the OMA scheme and the general OMA scheme are applicable. However, for the case with $K>2$, only the general OMA scheme is applicable. For the first $K-1$th devices, the distance of the $k$th device to the controller is set as $50\times k$ m, while the distance of the last device to the controller is set as $500$ m. The other parameters are the same as the previous figures. It is interesting to find that the decoding error probability achieved by the OMA scheme and the general OMA scheme is almost the same when $K=2$, which implies that the general OMA can achieve almost the globally optimal solution in this setup. However, the general OMA scheme has lower complexity than the OMA scheme. It is also noted from this figure that the decoding error probability achieved by the $K$th device increases when the number of total devices increases. This can be explained as follows. When the number of total devices increases, the total resource such as energy and channel blocklength allocated to the first $K-1$ devices will increase. Then, the left resource allocated for the $K$th device decreases, leading to its sworse decoding error probability performance.}

\vspace{-0.4cm}
\subsection{Network Availability  Performance (Channel Generation Times=1000)}
\vspace{-0.2cm}

In this subsection, the small-scale fading channel is taken into consideration in the channel gain, and we study the network availability  performance, which is defined as the ratio of the number of channel generations, where the decoding error probability
achieved by both devices is no larger than $10^{-9}$, to the total number of channel generations \cite{Schulz2017}. In the following simulations, the total number of channel generations is set as $1000$. The distances are set as $d_1=200$ m, $d_2=500$ m, and $d_3=300$ m, respectively.

Fig.~\ref{packetna} illustrates the network availability performance versus the packet size $D$ for all schemes. As expected, the network availability performance achieved by all schemes decreases with $D$. The relay-assisted transmission has the best network availability performance over the whole region of $D$. It is observed that when $D=100$ bits, the network availability percentage of the relay-assisted scheme and the C-NOMA scheme is almost the same, as high as 98\%. However, the performance gap of these two schemes increases rapidly with $D$ due to the shrinking feasible region of the C-NOMA scheme compared to the relay-assisted transmission. However, the network availability performance for both the OMA scheme and the NOMA scheme are lower than that of relay-assisted scheme and C-NOMA scheme, and the network availability percentage is as low as 87\% for NOMA scheme even when $D= 100$ bits.

\begin{figure}
\begin{minipage}[t]{0.5\linewidth}
\setcaptionwidth{0.95 \textwidth}
\captionsetup{font=small}
\centering
\includegraphics[width=3.1in]{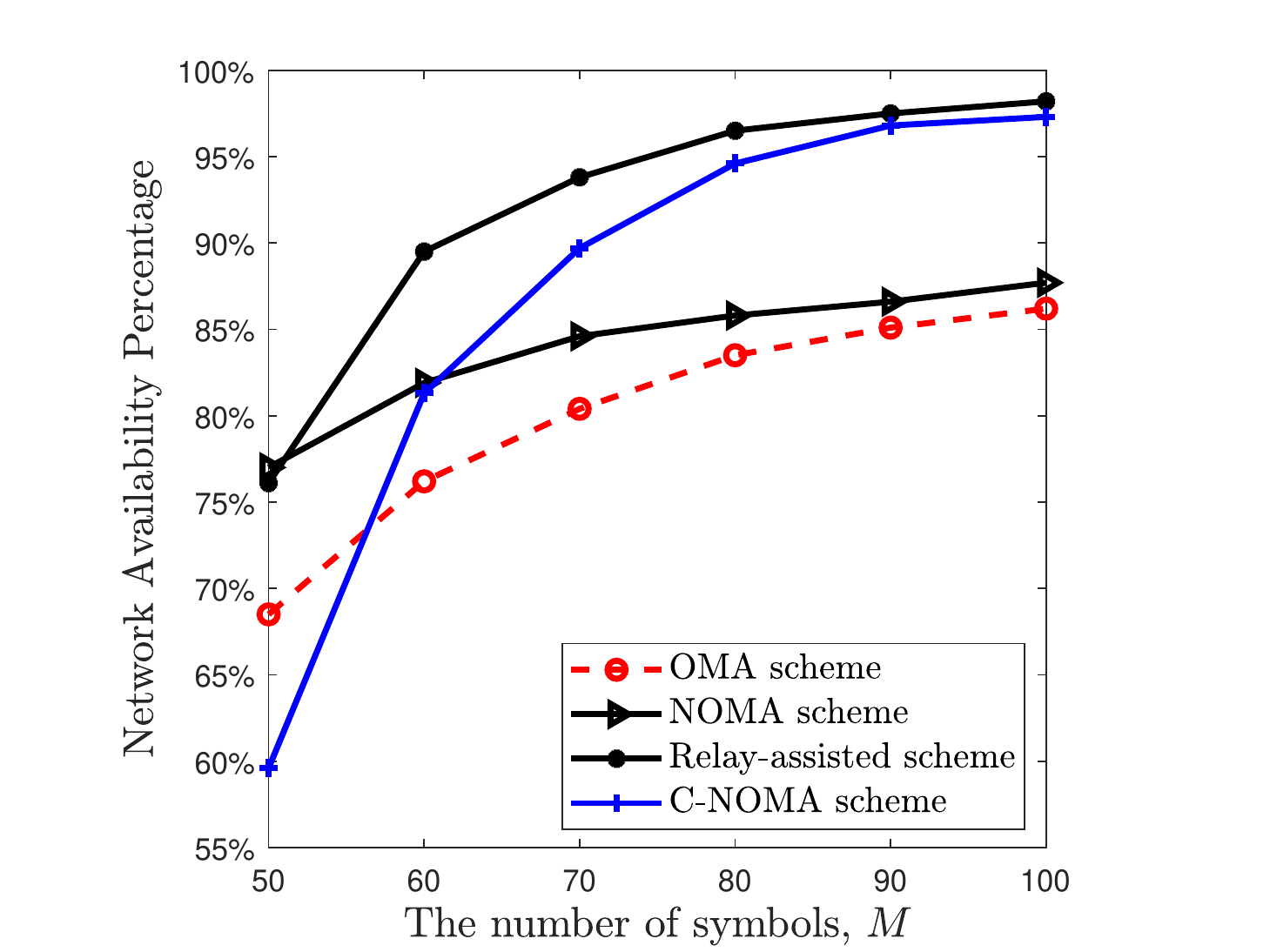}
\vspace{-0.3cm}
\caption{The network availability percentage versus the number of symbols $M$ under four schemes, when $\tilde{E}_{\rm{tot}}=5\times10^{-4}$ Joule, $D=100$ bits.}
\label{symbolsnetavail}
\vspace{-0.7cm}
\end{minipage}
\begin{minipage}[t]{0.5\linewidth}
\captionsetup{font=small}
\setcaptionwidth{0.95 \textwidth}
\centering
%  % Requires \usepackage{graphicx}
%\includegraphics[width=\textwidth]{./fig/Symbol.pdf}
\includegraphics[width=3.1in]{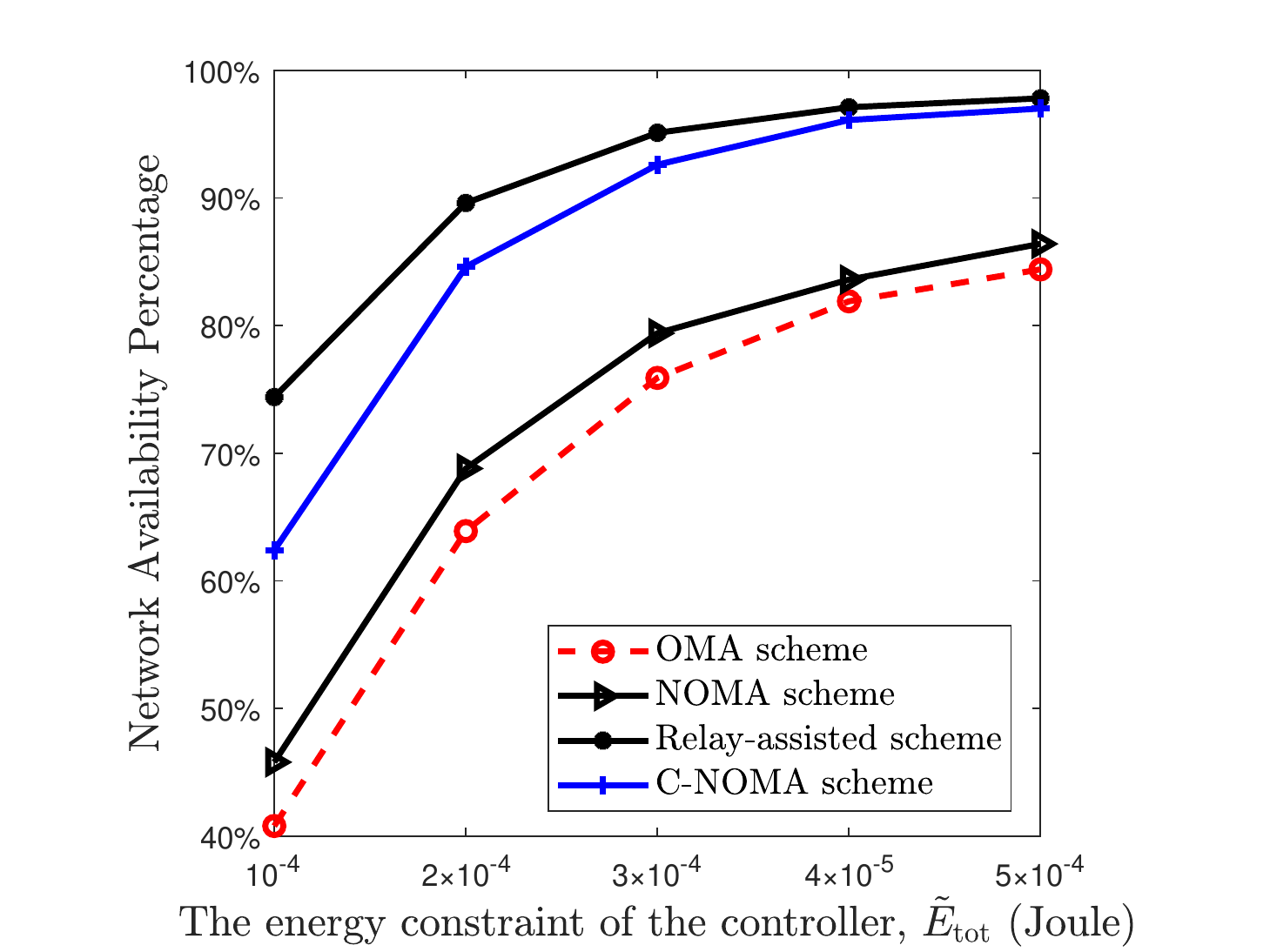}
\vspace{-0.3cm}
\caption{The network availability percentage versus energy limit  under four schemes, when $D=100$ bits, $M=100$ symbols.}
\label{energynet}
\vspace{-0.5cm}
\end{minipage}%
\vspace{-0.7cm}
\end{figure}

Fig.~\ref{symbolsnetavail} shows the network availability performance versus the number of symbols for four schemes. As expected, the network availability performance increases with $M$ for all schemes. The NOMA scheme performs slightly better than the relay scheme when $M=50$. It is interesting to note that the C-NOMA scheme has the worst performance when $M=50$, which means that this scheme is not a good option when there is stringent latency requirement. However, the network availability percentage of the C-NOMA increases rapidly with $M$, and finally converges to almost the same value as that of the relay-assisted scheme, that is equal to 97\% when $M=100$.  It is also noted that the OMA scheme converges to almost the same performance as that of the NOMA scheme, and is low (86\% when $M=100$). It is interesting to find that the network availability performance of all the schemes saturates in the high region of $M$, which indicates that the number of available blocklength is not necessary to be very large.  This can be explained by using the result in \cite{weiyang}: The dispersion of quasi-static fading channels converges to zero, which implies that the maximum achievable data rate converges quickly to the outage capacity.

Finally, Fig.~\ref{energynet} depicts the network availability performance versus the energy limit ${\tilde E_{{\rm{tot}}}}$ for all schemes. As expected, the network performance achieved by all schemes increases with ${\tilde E_{{\rm{tot}}}}$. It is also observed that relay-assisted scheme has the best network availability performance. However, the performance gain over the C-NOMA scheme decreases with ${\tilde E_{{\rm{tot}}}}$ and both curves coincide in the high regime of ${\tilde E_{{\rm{tot}}}}$, where both schemes can achieve the network availability percentage of 98\%. On the other hand, both the NOMA scheme and OMA scheme have very low network availability percentage, e.g., 86\% when ${\tilde E_{{\rm{tot}}}}=5 \times {10^{ - 4}}$ Joule. The performance gap between the relay-assisted scheme and NOMA is significant, up to 30\%.

\section{Conclusions}\label{conclusion}
This work studied the resource allocation of short packet transmission for mission-critical IoT to achieve low latency and high reliability under fundamental transmission schemes, which  include OMA, NOMA,  relay-assisted transmission and C-NOMA transmission. We formulated an optimization problem to minimize the decoding error probability for the actuator with lower channel gain while guaranteeing that the robot achieved a low error probability target. To facilitate the optimal design of the blocklength and power allocation, we derived the tight bounds on the blocklength and the transmit power for all schemes.  Simulation results demonstrated that relay-assisted transmission significantly outperforms the other schemes for most cases in terms of packet error probability as well as network availability percentage performance. It was also noted that the NOMA scheme performs well when the delay requirement is very stringent. For the C-NOMA and relay-assisted schemes, there exists one optimal transmission distance between the central controller and the robot. { We also observed that the general OMA scheme can achieve almost the same performance as the OMA scheme, while the former scheme has a lower complexity.}

 {Concerning our future work, we will consider a more general scenario with more than two devices for the other three schemes.}

\begin{appendices}
\section{Proof of Lemma 1}\label{lemma1}
We prove it by using contradiction. In the following, we first prove that constraint (\ref{rgegrtgue}) holds with equality at the optimum solution. The second one can be proved similarly.

Denote the optimal solution of Problem (\ref{initial-pro1}) as ${\bf{s}}^\star=\{m_1^\star, m_2^\star, {p_1^\star},{p_2^\star}\}$ and  the corresponding ${ \varepsilon }_1$ and ${ \varepsilon }_2$ are denoted as ${ \varepsilon }_1^\star$ and $ \varepsilon _2^\star$, respectively. Suppose that ${ \varepsilon }_1^\star$ is strictly smaller than $\varepsilon _1^{\max }$, i.e., ${ \varepsilon }_1^\star<\varepsilon _1^{\max}$. In Proposition 1 of \cite{xiaoyusun}, the author proved that $Q\left( {f\left( {{{\gamma_1}},{m_1},D} \right)} \right)$ monotonically  decreases with $\gamma_1$. Then, we can construct a new solution
${\bf{s}}^\#=\{m_1^\star, m_2^\star, {p_1^\#},{p_2^\#}\}$, where  ${p_1^\#}={p_1^\star}-\Delta p$ and ${p_2^\#}={p_2^\star}+\frac{{m_1^ \star \Delta p}}{{m_2^ \star }}$ with $\Delta p>0$.  It can be verified that the following equation holds,
\vspace{-0.3cm}
\begin{spacing}{1.0}
\begin{equation}\label{dewhufru}
\vspace{-0.1cm}
  m_1^ \star p_1^\#  + m_2^ \star p_2^\#=m_1^ \star p_1^ \star  + m_2^ \star p_2^ \star \le {E_{{\rm{tot}}}}.
\end{equation}
\end{spacing}
Hence, the new constructed solution ${\bf{s}}^\#$ still satisfies the energy constraint (\ref{fgrtgrtihre}). In addition, we can always find a proper positive $\Delta p$ such that the new ${ \varepsilon }_1^\#$ with the new solution ${\bf{s}}^\#$ is equal to $\varepsilon _1^{\max }$, i.e., ${ \varepsilon }_1^\#=\varepsilon _1^{\max }$, which satisfies constraint (\ref{rgegrtgue}). Hence, the new constructed solution ${\bf{s}}^\#$ is a feasible solution of Problem (\ref{initial-pro1}). Since $p_2^\#>p_2^\star$, we have ${ \varepsilon }_2^\#<{ \varepsilon }_2^\star$. This contradicts with the assumption that ${\bf{s}}^\star$ is an optimal solution. The same method is applicable to the proof of the second conclusion.

\vspace{-0.6cm}
\section{Proof of Theorem 1}\label{theorem1}
The first and second derivative of function $\tilde g({m_2})$ w.r.t. $m_2$ can be calculated as
\vspace{-0.4cm}
\begin{spacing}{1.1}
\begin{eqnarray}
\vspace{-0.4cm}
\tilde g'({m_2})& =& \frac{1}{{2\ln 2}}\frac{1}{{\sqrt m_2 }}\ln \left( {1 + \frac{{{E_2}{h_2}}}{{{m_2}}}} \right) - \frac{1}{{\ln 2}}\frac{1}{{\sqrt m_2 }}\frac{{{E_2}{h_2}}}{{{m_2} + {E_2}{h_2}}} + \frac{D}{2}m_2^{ - \frac{3}{2}}\nonumber\\
\tilde g''({m_2}) &=& \underbrace { - \frac{1}{4\ln2}\frac{{\ln \left( {1 + \frac{{{E_2}{h_2}}}{{{m_2}}}} \right)}}{{{m_2{\sqrt m_2}}}} + \frac{{{E_2}{h_2}}}{{{\ln 2{\sqrt m_2}{\left( {{m_2} + {E_2}{h_2}} \right)}^2}}}}_?\underbrace { - \frac{3}{4}Dm_2^{ - \frac{5}{2}}}_{ < 0}. \nonumber
\end{eqnarray}
\end{spacing}
Obviously, the last term of $g''({m_2})$ is negative, we only need to prove that the sum of the first two terms is negative under the condition of  $\frac{{{E_2}{h_2}}}{{M - {m_1}}} \ge e - 1$.

Since $m_{\rm{2}}^{{\rm{lb}}} \le {m_2} \le M - {m_1}$, we have
\vspace{-0.1cm}
\begin{spacing}{1.2}
\begin{equation}\label{wdwde}
  \frac{{{E_2}{h_2}}}{{{m_2}}} \ge \frac{{{E_2}{h_2}}}{{M - {m_1}}} \ge e - 1.
\end{equation}
\end{spacing}
Then, the following inequality follows:
\vspace{-0.1cm}
\begin{spacing}{1.2}
\begin{equation}\label{swqsed}
4 \le \left( {\frac{{{E_2}{h_2}}}{{{m_2}}} + 2 + \frac{{{m_2}}}{{{E_2}{h_2}}}} \right)\ln \left( {1 + \frac{{{E_2}{h_2}}}{{{m_2}}}} \right).
\end{equation}
\end{spacing}
By rearranging the terms of the above inequality, we can prove that the sum of the first two terms is negative, which completes the proof.

\vspace{-0.5cm}
{\section{Proof of Theorem 2}\label{theorem2}
\vspace{-0.2cm}

 We prove this theorem by using the method of contradiction. Denote the optimal $p_1$ of Problem (\ref{initial-pro2}) as  $p_1^\star$ and  the corresponding decoding error probability is given by ${\bar \varepsilon }_1(p_1^\star)$. Suppose that ${\bar \varepsilon }_1(p_1^\star)$ is strictly less than $\varepsilon _1^{\max }$, i.e., ${\bar \varepsilon }_1(p_1^\star)<\varepsilon _1^{\max }$. Since $\hat\varepsilon _1(p_1^{{\rm{lb}}}) > {\varepsilon _1}(p_1^{{\rm{lb}}})$, we have
\begin{equation}
{{\bar \varepsilon }_1}(p_1^{{\rm{lb}}}) = {\varepsilon _1}(p_1^{{\rm{lb}}}) + (\hat\varepsilon _1(p_1^{{\rm{lb}}}) - {\varepsilon _1}(p_1^{{\rm{lb}}}))\varepsilon _2^1(p_1^{{\rm{lb}}})= \varepsilon _1^{\max } + (\hat\varepsilon _1(p_1^{{\rm{lb}}}) - {\varepsilon _1}(p_1^{{\rm{lb}}}))\varepsilon _2^1(p_1^{{\rm{lb}}})>\varepsilon _1^{\max },
\end{equation}
where ${\varepsilon _1}(p_1^{{\rm{lb}}})=\varepsilon _1^{\max }$ is used in the second equality. As ${{\bar \varepsilon }_1}({p_1})$ is a continuous function,  there must exist a value $p_1^\&$ within the range of $p_1^{{\rm{lb}}}<p_1^\&<p_1^\star$ such that ${{\bar \varepsilon }_1}({p_1^\&})=\varepsilon _1^{\max }$. On the other hand, the objective value $\varepsilon _2(p_1)$ is a monotonically increasing function of $p_1$ since $p_2=E_{\rm{tot}}/m-p_1$. Hence, we have $\varepsilon _2(p_1^\&)<\varepsilon _2(p_1^\star)$,
which contradicts the assumption that  $p_1^\star$ is an optimal solution.}

\vspace{-0.3cm}
{\section{Proof of Theorem 3}\label{theorem3}
\vspace{-0.1cm}

We first prove its convexity. Define function
\begin{equation}\label{acfvv}
  J({m_k}) \buildrel \Delta \over = {m_k}{2^{\frac{D}{{{m_k}}} + \frac{{{A_k}}}{{\sqrt {{m_k}} }}}}.
\end{equation}
Then, $g\left( {{m_k}} \right)$ can be rewritten as $g\left( {{m_k}} \right)={{\left( {J({m_k}) - {m_k}} \right)} \mathord{\left/
 {\vphantom {{\left( {J({m_k}) - {m_k}} \right)} {{h_k}}}} \right.
 \kern-\nulldelimiterspace} {{h_k}}}$. Then, if $J({m_k})$ is convex, function $g\left( {{m_k}} \right)$ is also convex. Hence, in the following, we prove that $J({m_k})$ is a convex function. Define function ${\tilde J({m_k})}$ as
\begin{equation}\label{asdcfrgat}
\tilde J({m_k}) \buildrel \Delta \over = \ln \left( {J({m_k})} \right) = \ln ({m_k}) + \left( {\frac{D}{{{m_k}}} + \frac{{{A_k}}}{{\sqrt {{m_k}} }}} \right)\ln 2.
\end{equation}
The second-order derivative of $\tilde J({m_k})$ w.r.t. $m_k$ is given by
\begin{equation}\label{dewfr}
  \tilde J''({m_k}) = \frac{1}{{m_k^3}}\left( {2D\ln 2  - {m_k} + \frac{3}{4}{A_k}\sqrt {{m_k}}\ln 2 } \right).
\end{equation}
Note that the denominator of (\ref{dewfr}) is  a quadratic function of ${\sqrt {{m_k}} }$. Hence, if the inequality in (\ref{fdvtgxdtr}) is satisfied, $\tilde J''({m_k}) $ is always positive, which means $\tilde J({m_k})$ is a convex function of $m_k$. Since $J({m_k}) = {e^{\tilde J({m_k})}}$, according to the composition rule in \cite{boyd}, we can show that $J({m_k})$ is also a convex function. Hence, $g\left( {{m_k}} \right)$ is a convex function of $m_k$ when the inequality in (\ref{fdvtgxdtr}) is satisfied.

Now, we proceed to prove that $g\left( {{m_k}} \right)$ is a monotonically decreasing function of $m_k$. The first-order derivative of $g\left( {{m_k}} \right)$ w.r.t. $m_k$ is given by
\begin{equation}\label{dfrefe}
 g'\left( {{m_k}} \right) = \frac{1}{{{h_k}}}\left[ {{2^{\frac{D}{{{m_k}}} + \frac{{{A_k}}}{{\sqrt {{m_k}} }}}}\left( { - \frac{D}{{{m_k}}}\ln 2 - \frac{{\ln 2}}{2}\frac{{{A_k}}}{{\sqrt {{m_k}} }} + 1} \right) - 1} \right].
\end{equation}
Since $g\left( {{m_k}} \right)$ is a convex function, we have $g''\left( {{m_k}} \right)\ge 0$, which means $g'\left( {{m_k}} \right)$ is a monotonically increasing function. Hence, we have
\begin{equation}\label{dcefra}
g'\left( {{m_k}} \right) < g'\left( \infty  \right) = 0.
\end{equation}
Hence, $g\left( {{m_k}} \right)$ is a monotonically decreasing function of $m_k$ when the inequality in (\ref{fdvtgxdtr}) holds.}

\end{appendices}

\vspace{-0.1cm}
\bibliographystyle{IEEEtran}
 %argument is your BibTeX string definitions and bibliography database(s)
\bibliography{myre}
% that's all folks

\end{document}